\documentclass[12pt]{article}
\usepackage{epsfig,amssymb,amsmath,psfrag,graphicx,slashed}
\usepackage[ps2pdf]{hyperref} 
\def \bl  {\begin{align*}}
\def \el  {\end{align*}}

\def \be  {\begin{equation}}
\def \ee  {\end{equation}}
\def \ba  {\begin{eqnarray}}
\def \ea  {\end{eqnarray}}
\def \baa {\begin{eqnarray*}}
\def \eaa {\end{eqnarray*}}
\def \bb  {\begin {thebibliography} }
\def \eb  {\end{thebibliography}}
\def \lab #1 {\label{#1}}

\def \Tr {\mathop{\rm Tr}\nolimits}

\newcommand \vev [1] {\langle{#1}\rangle}

\newcommand{\ft}[2]{{\textstyle\frac{#1}{#2}}}

\newcommand{\nn}{\nonumber}
\newcommand{\cN}{{\cal N}}
\newcommand{\cO}{{\cal O}}
\newcommand{\p}[1]{(\ref{#1})}

\renewcommand{\title}[1]{\vbox{\center\LARGE{#1}}\vspace{5mm}}
\renewcommand{\author}[1]{\vbox{\center#1}\vspace{5mm}}
\newcommand{\address}[1]{\vbox{\center\em#1}}
\newcommand{\email}[1]{\vbox{\center\tt#1}\vspace{5mm}}
\newcommand{\eqn}[1]{(\ref{#1})}

\renewcommand{\=}{\mathrel{\phantom{=}}}

\begin{document}

\begin{titlepage}
\begin{center}
\vspace{5mm}
\hfill {\tt HU-EP-09/35}\\
\vspace{20mm}

\title{\sc Scattering into the fifth dimension \\of $\mathcal{N}=4$ super Yang-Mills}
\author{\large Luis F. Alday${}^{1}$,  Johannes Henn${}^{2}$, Jan Plef\/ka${}^{2}$ and Theodor Schuster${}^{2}$}
\address{${}^{1}$
School of Natural Sciences, Institute for Advanced Study \\
Princeton, NJ 08540, USA
\\[1cm]
${}^{2}$Institut f\"ur Physik, Humboldt-Universit\"at zu Berlin,\\
Newtonstra{\ss}e 15, D-12489 Berlin, Germany
}

\email{alday@ias.edu,\{henn,plefka,theodor\}@physik.hu-berlin.de}

\end{center}

\abstract{
\noindent
We study an alternative to dimensional regularisation of planar scattering
amplitudes in $\mathcal{N}=4$ super Yang-Mills theory by going to the Coulomb phase of the theory.
The infrared divergences are regulated by masses obtained from a Higgs mechanism,
allowing us to work in four dimensions.
The corresponding string theory set-up suggests that the
amplitudes have an exact dual conformal symmetry.
The latter acts on the kinematical variables of the
amplitudes as well as on the Higgs masses in an effectively five dimensional space. 
We confirm this expectation by an explicit calculation in the
gauge theory. A consequence of this exact
dual conformal symmetry is a significantly reduced set of scalar basis integrals that
are allowed to appear in an amplitude. For example, triangle sub-graphs are ruled out. We argue that the
study of exponentiation of amplitudes is simpler in the Higgsed theory because evanescent terms in
the mass regulator can be consistently dropped.
We illustrate this by showing the exponentiation of a four-point amplitude to two loops.
Finally, we also analytically compute the small mass expansion of a two-loop master integral with an internal mass.
\vfill
}

\end{titlepage}

\tableofcontents

\section{Introduction}

The maximally supersymmetric Yang-Mills theory ($\mathcal{N}=4$ SYM) in four dimensions
is special as it is the most symmetric version of a four dimensional gauge theory and possesses a host of
interesting features: It has a powerful quantum superconformal symmetry due to its vanishing
$\beta$-function, thus leaving the massless $U(N)$ theory controlled by only two tuneable parameters,
the number of colours $N$ and the
coupling constant $g_{\rm YM}$. Furthermore, highly nontrivial evidence has been accumulated
in favour of the AdS/CFT conjecture, claiming an exact duality
to the maximally  supersymmetric superstring theory on an $AdS_5\times S^5$ space-time
background \cite{AdS/CFT}.
In the planar 't Hooft limit of the ${\cal N}=4$ SYM model, where the interactions in the
dual string theory are absent, the gauge/string duality system displays fascinating integrable structures.
Prominently, the spectrum of anomalous dimensions of local operators is governed by an integrable
model \cite{AdSINT}, providing formulae valid to high loop orders or even at finite
't Hooft coupling $g^{2}_{\rm YM}\, N$  (see \cite{AdSrevs} for reviews).  \\

Similarly, the study of on-shell scattering amplitudes in the theory has seen substantial progress
in recent years. These amplitudes are also of phenomenological interest due to their relation to
QCD scattering amplitudes. Here the spinor helicity formalism \cite{spinhel} and the twistor space
approach \cite{Witten:2003nn} has initiated studies leading to important new insights.
As such, recursion relations
for tree-level amplitudes of ${\cal N}=4$ SYM have been established \cite{Britto}
and their on-shell superspace
formulation \cite{superrec} led to an explicit analytic solution of all tree-level amplitudes
\cite{Drummond:2008cr}, which includes all gluon trees in QCD.
At the loop level the development of generalised
unitarity techniques introduced in \cite{unitarity94} (see \cite{OSrev} for a review) was the key to very impressive
progress. For instance, these techniques enabled the computation of four-gluon amplitudes up to four loops
\cite{Bern:2006ew} and six-gluon amplitudes up to two loops \cite{Bern:2008ap}. \\

Based on an iterative structure \cite{Anastasiou:2003kj} found at lower loop levels,
Bern, Dixon and Smirnov (BDS)  \cite{Bern:2005iz} conjectured an all-loop form of the
maximally helicity violating (MHV) amplitudes. By now, this ansatz is believed to be correct for four and five gluon amplitudes
(but known to fail for more than five particles \cite{Bern:2008ap}, see also \cite{Alday:2007he,Bartels:2008ce,DHKS6pt}). 
The correctness of the BDS ansatz for four and five gluons stems from a novel
hidden symmetry of the planar theory, dual conformal symmetry. Hints for this symmetry first appeared in \cite{Drummond:2006rz}, and then
independently in \cite{Alday:2007hr}.
Since then, it has been developed \cite{Drummond:2007aua,Drummond:2007cf, Drummond:2007au} and, importantly, it was discovered that it extends to a 
dual superconformal symmetry \cite{Drummond:2008vq}.
In particular, 
the latter
is a 
symmetry of all tree-level amplitudes, 
as shown in \cite{Drummond:2008vq,Brandhuber:2008pf,Drummond:2008cr}.\footnote{See also the recent papers \cite{Bargheer:2009qu,Korchemsky:2009hm}.}\\

The dual conformal symmetry can be understood through the string theory description
of scattering amplitudes at strong coupling \cite{Alday:2007hr,Alday:2007he}, which identifies
the scattering amplitude calculation with a Wilson loop computation in a T-dual $AdS$ space.
Dual conformal symmetry is then interpreted as the usual conformal symmetry of the dual Wilson loop. 
{Furthermore, the dual superconformal symmetry of \cite{Drummond:2008vq} can also be seen from the string theory perspective, through a novel fermionic T-duality \cite{Berkovits:2008ic,Beisert:2008iq}.}
 \\

Quite remarkably the scattering amplitude/Wilson loop relation extends all the way down to weak
coupling \cite{Drummond:2007aua, Brandhuber:2007yx, Drummond:2007cf} (for reviews see \cite{AWLreviews}).
The dual conformal symmetry is anomalous at loop level due to
ultraviolet divergences associated with cusps of the Wilson loops
(which are related to the infrared divergences of the scattering amplitudes \cite{Ivanov:1985np}).
However, the breaking of dual conformal symmetry is under full control and can be written in terms
of all-order anomalous Ward identities derived in \cite{Drummond:2007cf,Drummond:2007au} 
{(see also \cite{Komargodski:2008wa,AWLreviews})}.
In particular, the latter determine the finite part of the Wilson loops for four and five cusps to
be of the form of the BDS ansatz, to all orders in the coupling constant.
The dual conformal anomaly is proportional to the anomalous dimension
of a light-like Wilson loop cusp \cite{WLren},
a universal function in turn 
{conjectured to be exactly known}
as a key outcome of the
above mentioned AdS/CFT integrability investigations \cite{Beisert:2006ez}.
The existence of two copies of the superconformal symmetry algebra is a hallmark of integrability
\cite{Ricci:2007eq,Berkovits:2008ic,Beisert:2008iq}, as their closure results in an
infinite dimensional symmetry algebra of Yangian structure under which the tree-level amplitudes
are invariant \cite{Yangian}. At the loop-level the status of the Yangian symmetry is unclear at present:
The IR divergences destroy both the standard and dual superconformal symmetries. However, while the
breaking of the dual conformal symmetry can be controlled, similar control does not (yet) exist for the standard conformal symmetry.
A key issue here, and one of the motivations for this work, is clearly the regularisation prescription and its
behaviour under the conformal symmetry transformations.\\

The most widespread regularisation is certainly dimensional regularisation, or
rather dimensional reduction in order to preserve supersymmetry.
This method is very well developed. An inconvenience
of this regularisation is that when computing for example the logarithm of an
amplitude, as suggested by the form of infrared divergences and the BDS ansatz,
there is an interference between poles in the dimensional regulator $\epsilon$ and
evanescent terms in $\epsilon$ coming from lower-loop amplitudes. As a result,
one has to compute these higher order $\epsilon$ terms in the lower-loop amplitudes.\\

An idea to circumvent this problem was proposed in \cite{Drummond:2007aua}, where an off-shell
regulator was used. Divergences in this regulator would take the form of logarithms,
and therefore, the above interference could not take place, at least to a given order
in the coupling constant. Also, one could have hoped that this regulator is more suited
to expose dual conformal symmetry.
Unfortunately, the use of an off-shell regulator leads to
other problems such as the lack of manifest gauge invariance. \\

However, there is another regularisation motivated naturally by the dual string picture
that is somewhat similar in spirit but different
from the off-shell regularisation, which we shall employ. This regularisation was discussed in \cite{Alday:2007hr,Kawai:2007eg, Schabinger:2008ah,McGreevy:2008zy,Berkovits:2008ic} and consists
in turning on a vacuum expectation value for one of the scalars in $\mathcal{N}=4$ SYM.
Specifically, one takes a $U(N+M)$ theory and applies the Higgs mechanism to break the symmetry to
$U(N) \times U(M)$. {Then, one considers the scattering of the $U(M)$ fields, which lead to massive
propagators in the loops.
In the $N \gg M$ limit \footnote{Note that we do not set $M=1$ because we want to be able to define
a colour ordering for the outer legs.}, the following picture emerges:
If we use a double-line notation, then the $U(M)$ lines will be on the outside of the diagram,
while in the interior we will have $U(N)$ lines only. Hence the massive particles will 
flow around the outer line of the diagram, and thereby regulate the infrared divergences. 
Hence in the planar, large $N$ limit, one can consider scattering processes in the Higgsed theory
that are regulated by the Higgs mass and therefore can be defined in four dimensions. 
We expect this regularisation to work to all orders in the coupling constant. }\\

\begin{figure}
\label{fone}
\psfrag{ads5}[cc][cc]{AdS${}_5$}
\psfrag{N}[ll][ll]{$N$ D3-branes}
\psfrag{M}[ll][ll]{$M$ D3-branes}
\psfrag{z0}[rr][rr]{$z=0$}
\psfrag{zi}[rr][rr]{$z_{i}=1/m_{i}$}
\psfrag{aba}[cc][cc]{(a)}
\psfrag{p1}[cc][cc]{$p_{2}$}
\psfrag{p2}[cc][cc]{$p_{3}$}
\psfrag{p3}[cc][cc]{$p_{4}$}
\psfrag{p4}[cc][cc]{$p_{1}$}
\psfrag{i1}[cc][cc]{\tiny $i_{2}$}
\psfrag{i2}[cc][cc]{\tiny $i_{3}$}
\psfrag{i3}[cc][cc]{\tiny$i_{4}$}
\psfrag{i4}[cc][cc]{\tiny$i_{1}$}
\psfrag{n}[cc][cc]{\tiny$j$}
\psfrag{m}[cc][cc]{\tiny$k$}
\psfrag{box2aA}[cc][cc]{(b)}
 \centerline{
{\epsfysize6cm
\epsfbox{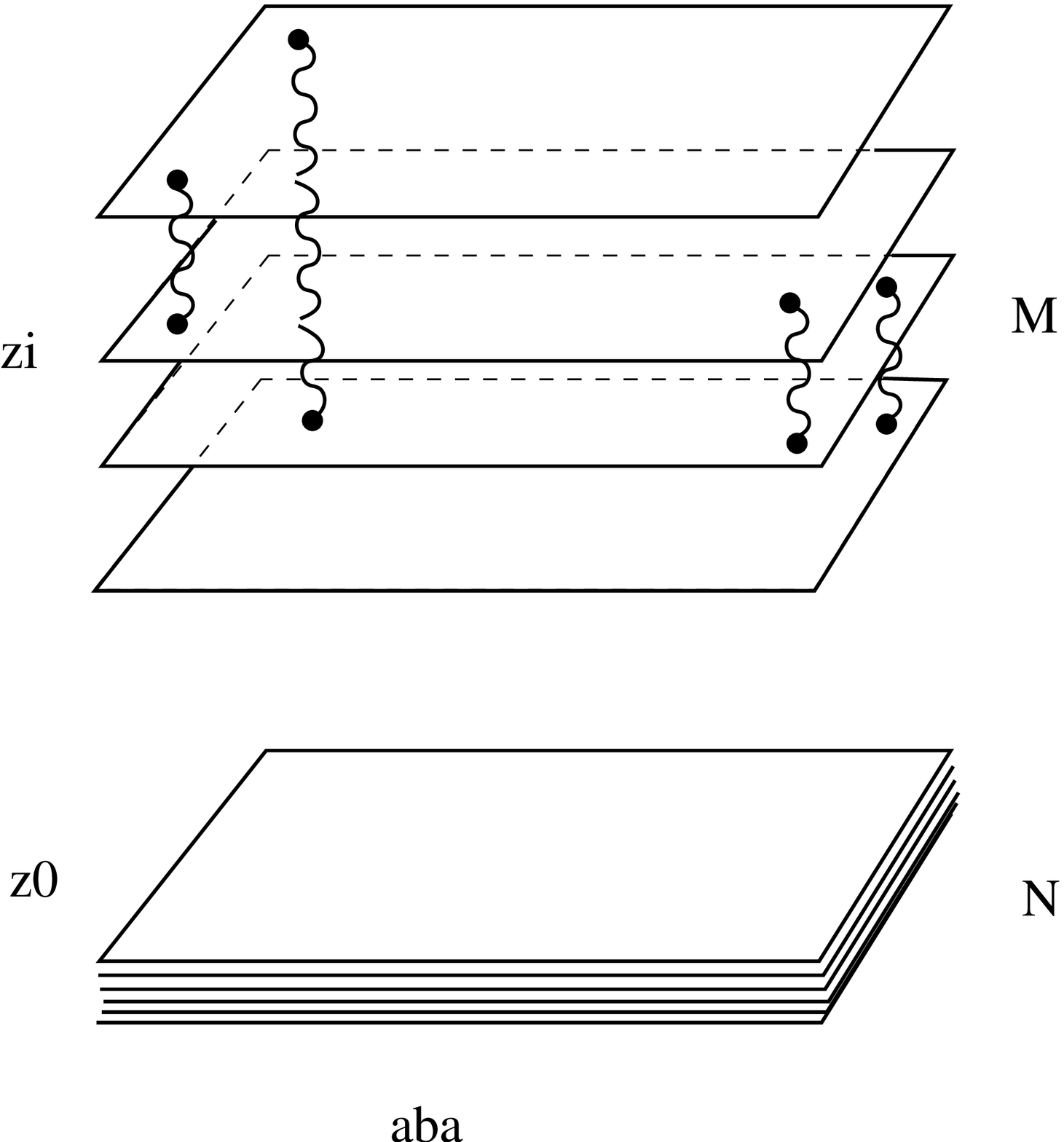}}
\hspace{3cm}
{\epsfysize5cm
\epsfbox{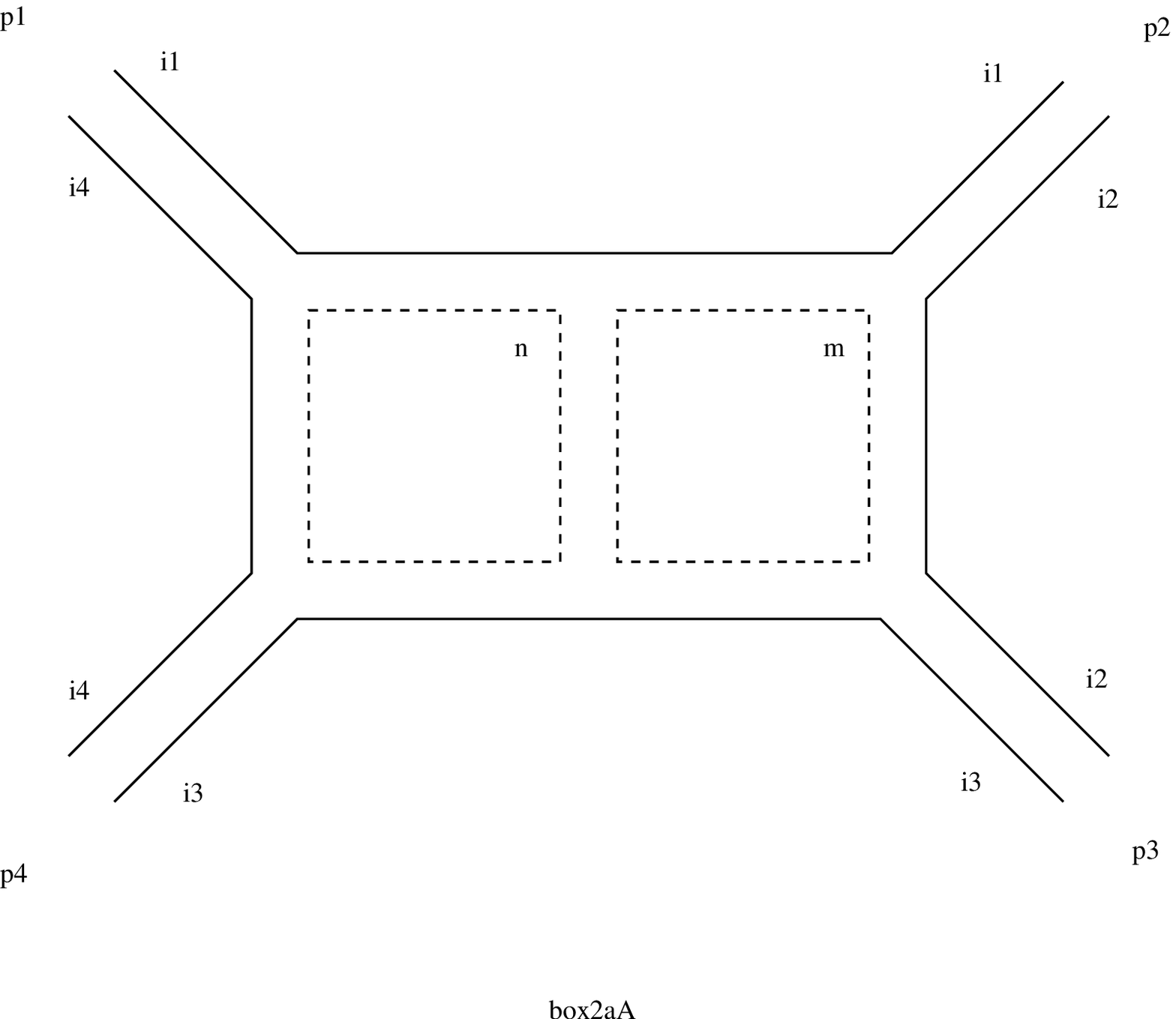}}
\vspace{0.5cm}}

\caption{\small
(a) String theory description for the scattering of $M$ gluons in the large $N$ limit.
Putting the $M$ D3-branes at different positions $z_{i}\neq 0$ serves as a regulator
and also allows us to exhibit dual conformal symmetry. (b) Gauge theory analogue of (a): a generic
scattering amplitude at large $N$ (here: a sample two-loop diagram).}
\end{figure}

Importantly, we can improve this set-up by allowing for different Higgs masses, breaking the
$U(N+M)$ gauge symmetry down to $U(N)\times U(1)^{M}$. In the dual string picture this amounts to moving
$M$ D3-branes away from the $N$ parallel D3-branes and also separating these $M$ distinct branes
from one another. One then has ``light'' gauge fields corresponding to strings stretching
between the $M$ separated D3-branes, which are our external scattering states. Then there are
the ``heavy'' gauge fields corresponding to the strings stretching between the coincident $N$
D3-branes and one of the $M$ branes. These are the massive particles running on the outer line
of the diagrams, see figure \ref{fone}. In doing so, we argue that dual conformal symmetry,
suitably extended to act on the Higgs masses as well, is an exact, i.e.~unbroken, symmetry of
the scattering amplitudes.\\

This exact symmetry has very profound consequences. It was already noticed
in \cite{Drummond:2006rz} that the integrals contributing to loop amplitudes in $\mathcal{N}=4$ SYM
have very special properties under dual conformal transformations, but this observation
was somewhat obscured by the infrared regulator. With our infrared regularisation,
the dual conformal symmetry is exact and hence so is the symmetry of the integrals.
Therefore, the loop integrals appearing in our regularisation will have an exact
dual conformal symmetry. This observation severely restricts the class of integrals
allowed to appear in an amplitude. As a simple application, triangle sub-graphs are
immediately excluded.\\

The alert reader might wonder whether computing a scattering amplitude with several,
distinct Higgs masses might not be hopelessly complicated. In fact, this is not the case.
The different masses are crucial for the exact dual conformal symmetry to work. However,
once we have used this symmetry in order to restrict the number of basis loop integrals, we
can set all Higgs masses equal and think about the common mass as a regulator. As we will
show in several examples, computing the small mass expansion in this regulator is
particularly simple. In fact, to two loops, only very simple (two-) and (one-)fold Mellin-Barnes integrals
are needed.\\

The reader may be worried that the infrared regulator we propose is not complete, i.e.~that
one might still find infrared divergences at some higher loop order from massless subgraphs. Infrared divergences come from regions of
the integration space where the loop momentum is soft and/or collinear to some external
momentum. At low loop level, we will see explicitly that the massive particles flowing around the outer line of the diagrams regulate these potential divergences. 
At higher loop order diagrams with massless subgraphs may occur, and while we do not have a formal 
proof, we do expect that also such diagrams are finite in our setup
\footnote{We are grateful to G. Korchemsky and L. Dixon for discussions of this point.}. 
An argument in favour of this
is that from the strong coupling string perspective there is no divergence.\\

This paper is organised as follows. In section two we describe scattering amplitudes from the string theory perspective in the above mentioned regularisation and argue, in agreement with \cite{Berkovits:2008ic}, that the amplitudes are expected to possess dual conformal invariance. In section three we consider the analogous regularisation in perturbation theory. In particular, we consider the case of the four point amplitude up to two loops and show that the expectations from the strong coupling side are indeed fulfilled. Furthermore, we show that exponentiation holds for this case. Finally, we present our conclusions and an outlook, referring to the appendices many technical details relevant to the body of the paper.

\section{String theory}
\label{sect-strings}

In this section we analyze the above mentioned scattering amplitudes from the string theory picture, which is the appropriate description around the regime of strong coupling. If one focuses on planar amplitudes, the appropriate world-sheet has the topology of
a disk, with vertex operator insertions at the boundary corresponding to the external states undergoing the scattering.\\

On the string side, the regularisation to be considered in this paper is quite natural and corresponds to introducing $M$ $D3-$branes in the background $AdS_5 \times S^5$. To be more precise, if we write the $AdS_5$ metric in Poincar\'e coordinates $ds^2=\frac{1}{z^{2}}(dy^2_{3,1}+dz^2)$,  then the $M$ branes are sitting at
the positions $z_i=1/m_i$ ($i=1,\ldots, M$)
and extend along the $y_{3,1}$ directions. The asymptotic states to be scattered are the open strings between a pair of consecutive  $D3-$branes, for instance at $z_i$ and $z_{i+1}$. These open strings represent the gluons.\\

\begin{figure}
 \centerline{{\epsfysize5.5cm
\epsfbox{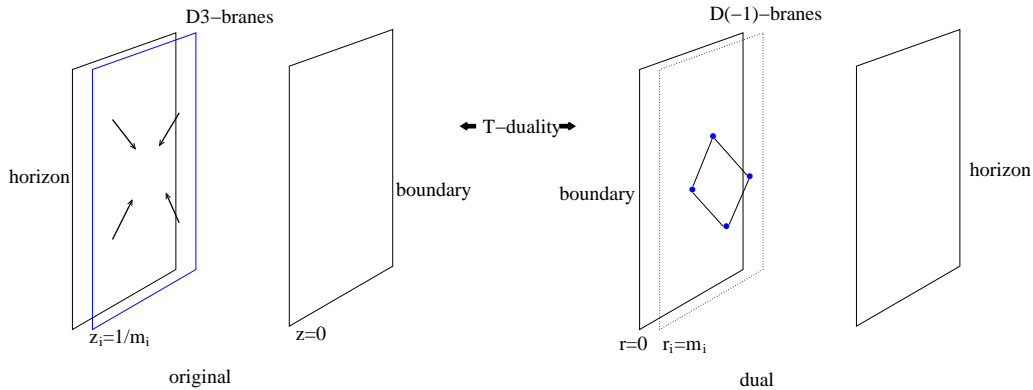}}}  \caption{\small
Original (left) and dual (right) pictures of a scattering amplitude. On the original picture the open strings end at $D3-$branes located at $z_i=m_i$. In the dual picture we have open strings stretched between $D-$instantons separated  by a light-like distance.}
\label{ordu}
\end{figure}

As argued in \cite{Alday:2007hr,Berkovits:2008ic}, to which we refer the reader for more details, it is convenient to perform four T-dualities
in the $y_{3+1}$ directions, followed by a change of coordinates $r=\frac{1}{z}$ (we are setting the $AdS$ radius to one for convenience). After this, we end up with a dual $AdS$ metric
\begin{equation}
ds^2=\frac{dx^2_{3,1}+dr^2}{r^2} \, .
\end{equation}
As $T-$duality interchanges Dirichlet and Neumann boundary conditions, the $D3-$branes become $D(-1)$-branes, or D-instantons. Each of these instantons is located at a fixed position in the $x_{3,1}$ coordinates and sits at $r_i=m_i$. The open strings are then stretching between consecutive D-instantons and the rules of
T-duality fix the distance between these instantons to be proportional to the momentum of the original external state that the open string represented, see figure \ref{ordu}.

If the D-instantons are away from the boundary, namely, $m_i>0$, the amplitude is finite.
On the other hand, it can only depend on the covariant $AdS$ distances between the D-instantons
(furthermore, at strong coupling, or when considering Wilson loops, the amplitude does not depend on the details of the inserted states). On the other hand, on dimensional grounds, we can only have dependence on ratios of
these distances.\\

The dual conformal symmetry, being the conformal symmetry in the T-dual space,
now acts in the above system by changing the location of the $D-$instantons $(r,x_{3,1}) \rightarrow (r',x_{3,1}')$. For instance, one can consider special conformal transformations, in which case one
has
\begin{align}
\label{confAdS}
r'&=\frac{r}{1+2\, x\cdot\beta+\beta^2\, (r^2+x^2)},\nonumber \\
(x')^\mu&=\frac{x^\mu+(r^2+x^2)\, \beta^\mu}{1+2\, x\cdot\beta+\beta^2(r^2+x^2)}
\end{align}
Since the amplitude is regularised, hence finite, and since the system possesses dual conformal symmetry, the amplitude should be invariant under these transformations, at least when not taking into account the contribution from the polarisation of the external states. \footnote{In other words, when considering the ``Wilson loop'' computation.}

This symmetry can be easily checked at strong coupling. In such a regime the amplitude does not depend on the details of the external states and is dominated by a saddle point of the classical action, whose lagrangian, in conformal gauge, reads
\begin{eqnarray}
{\cal L}=\frac{\partial_i r \partial_i r+\eta_{\mu \nu} \partial_i x^\mu \partial_i x^\nu}{r^2} \, .
\end{eqnarray}
One can check that (\ref{confAdS}) maps solutions of the equations of motion into solutions, keeping the Lagrangian invariant. Furthermore, the transformations are such that the boundary conditions are still the boundary conditions of a scattering problem, see discussion below. Hence the amplitude is invariant.\\

Unfortunately, it is hard to find classical solutions with boundary conditions at $r>0$ even for the
four cusp situation. However, the single cusp solution can be found in terms of a perturbation series, which
we derive in appendix \ref{stringapp}. This is the conformal gauge version of the Nambu-Goto solution found in \cite{Berkovits:2008ic}
and should describe the limiting form of a generic scattering string world-sheet when approaching any of the cusps.\\

Even though the full solution is not known, the single cusp solution allows one to extract the form of the cusp anomalous dimension
at strong coupling in this regularisation. We indeed find
\begin{align}
\lim_{\lambda\to\infty}\Gamma_{\text{cusp}}&= \frac{\sqrt{\lambda}}{\pi}\, , &\text{where}&&
\lambda&=g^{2}_{\rm YM} N \, ,
\end{align}
in agreement with the well known result.\\

The statement of invariance of the amplitudes under $SO(2,4)$ transformations can also be written
in a infinitesimal form (see appendix
\ref{AdS5app} for a derivation of the infinitesimal generators from the $AdS_{5}$ isometries). 
The relevant dual dilatation and special conformal generators take the form
\begin{align}\label{poincare-trans}
\hat D  &= r \partial_{r} + x^{\mu} \partial_{\mu}\,, \\
\hat K_{\mu} &= 2 \, x_{\mu} ( x_{\nu} \partial^{\nu} + r \partial_{r}) - (x^2 +r^2)\partial_{\mu}\,.
\end{align}

Now, we are interested in computing the classical string action $S$ for a world-sheet with suitable boundary conditions.
The action will be invariant under these transformations, but the boundary conditions might change, see below. We do not need to worry about a regulator as
long as the world-sheet does not end on the boundary at $r=0$. However, we should have boundary
conditions that transform nicely under (\ref{poincare-trans}).
For $r=0$, the appropriate boundary contour on which the world-sheet should
end is a polygon with $(x_{i} - x_{i+1})^2 = 0$. Importantly, such a light-like polygon is mapped into another light-like polygon. For $r \neq 0$, we see that the conditions
$(x_{i} - x_{i+1})^2 + (r_{i} - r_{i+1})^2 =0$ are similarly preserved by (\ref{poincare-trans}).
Let us denote the contour formed by the $M$ points $\{ x^{\mu}_{i},r_{i} \}$ by $C$.
Then, doing infinitesimal transformations we find that indeed
\begin{equation}\label{WLinv}
\hat{K}_{\mu} S(C) = 0\,.
\end{equation}
Where $K=\sum K_i$ and $K_i$ acts on the coordinates of the $i$th $D-$instanton. We stress that in order to write (\ref{WLinv}) we need to consider the amplitude for the case in which the $D-$instantons are at different radial distances $r_i$. On the other hand, even if we started with a configuration in which all the radial distances are the same, then a general dual conformal transformation would make them different.\\

The argument above only depends on using classical string theory, so it should be valid for the planar theory at large $\sqrt{\lambda}$. If we are interested in computing the Wilson loop expectation value, considering quantum fluctuations about such a minimal surface will not change the boundary conditions of the fields. Hence, we expect the dual conformal
symmetry to prevail to all orders in a $1/\sqrt{\lambda}$ expansion in the planar theory, as argued by \cite{Berkovits:2008ic}. \\

In order to find the same constraint on scattering amplitudes, one should understand how to introduce the dependence on the helicity of the external states. However,
it seems reasonable to assume that a formula very similar to (\ref{WLinv}) holds for scattering amplitudes as well. In the next section we will  indeed identify a class of scalar amplitudes
which for four particle scattering exhibit a parallel expression to \eqn{WLinv}, see
\eqn{exact-dualconformal1loop} and \eqn{exact-dualconformal} at one-loop order. 
We take this as an indication that this exact dual conformal symmetry is present from weak to
strong coupling.

\section{Gauge theory}
\label{sect-gauge}

\subsection{Higgsing \texorpdfstring{$\cN=4$}{N=4} super Yang Mills}
\label{sect-higgs}

Let us now work out the spontaneous symmetry breaking of $\cN=4$ SYM in more detail.
We consider the breaking of $U(N+M)\to U(N)\times U(1)^{M}$.
The component field spectrum consists of the vectors $A_{\mu}$, the six scalars $\Phi_{I}$ and the
ten dimensional Majorana-Weyl spinors $\Psi$ governed by the action
\be
{\hat S}^{U(N+M)}_{\cN=4}= \int d^{4}x\, \Tr \Bigl (
-\ft{1}{4}\, {\hat F}_{\mu\nu}^{2} -\ft{1}{2}(D_{\mu}{\hat \Phi}_{I})^{2} + \ft{g^2}{4}\, [{\hat \Phi}_{I},{\hat\Phi}_{J}]^{2} +\ft{i}{2}\,\hat{\overline{\Psi}}\,\Gamma^{\mu}\,D_{\mu}\hat{\Psi} +\ft{g}{2}\hat{\overline{\Psi}}\,\Gamma^{I}[{\hat \Phi_{I}},\hat\Psi]\, \Bigr ) \, ,
\ee
where $D_{\mu}=\partial_{\mu} -i g [A_{\mu}, \,\,\, ]$. All fields are hermitian matrices, which we decompose into blocks as
\begin{align}
{\hat A}_{\mu}&=
\begin{pmatrix} 
      (A_{\mu})_{ab} &  (A_{\mu})_{aj} \\
       (A_{\mu})_{ia} &  (A_{\mu})_{ij} \\
   \end{pmatrix} \, ,\qquad
 {\hat\Phi}_{I}=
   \begin{pmatrix} 
      (\Phi_{I})_{ab} &  (\Phi_{I})_{aj} \\
       (\Phi_{I})_{ia} &  \delta_{I9}\, \ft{m_{i}}{g}\,\delta_{ij}+(\Phi_{I})_{ij} \\
   \end{pmatrix} \, ,\qquad
 {\hat\Psi}=
   \begin{pmatrix} 
      (\Psi)_{ab} &  (\Psi)_{aj} \\
       (\Psi)_{ia} &  (\Psi)_{ij} \\
   \end{pmatrix} \, , \nn\\
   & a,b=1,\ldots, N\, ,i,j=N+1,\ldots, N+M\, ,
\end{align}
thereby turning on a vacuum expectation value (VEV) for the scalars ${\hat\Phi}_{I}=\delta_{I9}\, \vev{\Phi_{9}}+\Phi_{I}$
in the $I=9$ direction.
This shift introduces terms of linear and quadratic order in $m_{i}$
\begin{align}
{\hat S}_{\cN=4}={S}_{\cN=4} + \int d^{4}x\,  \Tr &\Bigl ( i g\, D_{\mu}\Phi_{9}
\,[A_{\mu},\vev{\Phi_{9}}] + \tfrac{g^2}{2}[A_{\mu},\vev{\Phi_{9}}\,]^{2}+ \tfrac{g^2}{2}[\Phi_{I'},\vev{\Phi_{9}}\,]^{2}
\nn\\ &
+ g^{2}\,[\vev{\Phi_{9}},\Phi_{J'}]\, [\Phi_{9},\Phi_{J'}] +\tfrac{g}{2}\overline{\Psi}\Gamma^{9}\, [\vev{\Phi_{9}},\Psi]
\,\Bigr ) \, ,
\label{shiftedSN4}
\end{align}
where $I',J'=4,\ldots, 8$.
We proceed by adding a $R_{\xi}$ gauge fixing term $-\frac{1}{2}\,\Tr(G^{2})$ with
\be
G= \frac{1}{\sqrt{\xi}}\, \Bigl [\partial_{\mu}A^{\mu}-ig\,\xi\, [\vev{\Phi_{9}}, \Phi_{9}]\,\Bigr ]
\, ,
\ee
and the appropriate ghost term
\begin{equation}
 {\cal L}_{\text{ghost}}=\Tr\left\{\bar c (\partial^\mu D_\mu c-g^2\xi[\vev{\Phi_9},[\Phi_9+\vev{\Phi_9},c]])\right\}\,.
\end{equation}
The gauge fixing term  $-\frac{1}{2}\,\Tr(G^{2})$
cancels the unwanted scalar-vector mixing first term in \p{shiftedSN4} and gives a gauge
parameter $\xi$ dependent mass term for $\Phi_{9}$ and $c$. We specialise to the choice $\xi=1$ to
obtain identical propagators for vectors and scalars.

The Higgsing adds mass terms and novel cubic interaction terms for the bosonic fields coupling to
$\Phi_{9}$, explicitly $\hat S_{\cN=4}$ of \p{shiftedSN4} now contains the quadratic terms
$(A_M:=(A_\mu,\Phi_I))$
\begin{align}
\hat S_{\cN=4}\Bigr|_{\text{quad.}} &=\int d^{4}x\,\Bigl\{{}-\tfrac{1}{2}
\Tr(\partial_{\mu} A_{M}\,\partial^{\mu}A^{M}) - \tfrac{1}{2}(m_{i}-m_{j})^{2}\, (A_M)_{ij}\, (A^{M})_{ji}
- m_i^2\, (A_M)_{ia}\,(A^{M})_{ai} \nn\\
&\=\phantom{\int d^{4}x\,\Bigl \{}+\tfrac{i}{2}\Tr(\overline{\Psi}\,\Gamma^\mu\,\partial_\mu\Psi) - \tfrac{1}{2}(m_{i}-m_{j})\,\overline{\Psi}_{ij}\Gamma^{9}\Psi_{ji}
+\tfrac{1}{2}m_{i}\,(\overline{\Psi}_{ai}\Gamma^{9}\Psi_{ia}-\overline{\Psi}_{ia}\Gamma^{9}\Psi_{ai})
\,\Bigr \}\,,
\end{align}
i.e.~we have the `light' fields $\cO_{ij}$ ($i\neq j$) with masses $(m_i-m_j)$, where $\cO$ denotes
a generic parton $\{A_{\mu}, \Phi_{I},\Psi\}$, and the heavy fields $\cO_{ia}$ of mass
$m_i$. The $\cO_{ab}$ and $\cO_{ii}$ remain
massless. Furthermore we pick up new cubic bosonic interaction terms proportional to $m_{i}$
\begin{align}
\hat S_{\cN=4}\Bigr|_{\cO(g\,m_{i})}&=g\,\int d^{4}x\,\Bigl \{
\, m_i\, ([\Phi_{9},A^{\mu}]\,A_\mu\,)_{ii} -m_i(A_\mu\,[\Phi_{9},A^{\mu}]\,)_{ii}\notag\\
&\=\phantom{\int d^{4}x\,\Bigl \{}+m_i\, ( [\Phi_{9},\Phi_{I'}]\,\Phi_{I'}\,)_{ii}-\, m_i\, (\Phi_{I'}\, [\Phi_{9},\Phi_{I'}]\,)_{ii}
\, \Bigr \}\, , \label{cubicmassV}
\end{align}
where the not-spelled out matrix index sums run over the full $N+M$ range.
Furthermore the ghosts $c$ and $\bar c$ will also receive a mass term and a new $\bar c\, c\, \Phi_{9}$ interaction term.
\\

Given this we note the following: If we use the VEVs $m_{i}$ as an IR regulator and
consider the scattering of colour ordered light gluons $(A_\mu)_{ij}$  (or scalars $(\Phi_{I})_{ij}$)
with $i\neq j$ along with the large $N$ 't Hooft limit, a one-loop computation of an $n$-particle
scattering process will involve precisely the same Feynman diagrams as in the $m_{i}=0$ case
but with massive $\cO_{ia}$ propagators. In particular the box integral will be that
of figure \ref{ftwo} (a).
In addition we will have new Feynman graphs involving the new $\cO(m_{i})$ 3-point
vertices \p{cubicmassV}. We will see in the next subsections that the new vertices
are engineered in precisely such a way that the amplitudes respect the dual conformal symmetry.\\

\subsection{One loop test of dual conformal symmetry}
\label{S-one-loop}

Here we want to investigate whether a perturbative calculation in the Higgsed version of
$\mathcal{N}=4$ SYM has the dual conformal symmetry discussed in section \ref{sect-strings}.
We choose a scattering amplitude of four scalars and compute it to one-loop level.
Specifically, we consider the colour-ordered amplitude
\begin{equation}\label{scalar4ptamplitude}
A_{4} = \langle \Phi_{4}(p_{1}) \, \Phi_{5}(p_{2}) \, \Phi_{4}(p_{3}) \,\Phi_{5}(p_{4})  \rangle\,,
\end{equation}
which is related to the leading colour contribution of the four scalar scattering amplitude by
\begin{equation}
{\mathcal A}_4=\sum_{\sigma\in S_4/Z_4}\delta_{i_{\sigma{(1)}}}^{j_{\sigma(1)}}\delta_{i_{\sigma(2)}}^{j_{\sigma(2)}}\delta_{i_{\sigma(3)}}^{j_{\sigma(3)}}\delta_{i_{\sigma(4)}}^{j_{\sigma(4)}}A_4(\sigma(1),\sigma(2),\sigma(3),\sigma(4))\,.
\end{equation}
Here $(i_{1},j_{1}),\ldots, (i_{4},j_{4})$ are the $U(M)$ matrix indices of the four scattered scalars, and
$\sigma$ stands for non-cyclic permutations of the set \{1,\ldots, 4\}.
The flavour choice of the scalars in (\ref{scalar4ptamplitude}) was made in such a way
that a proliferation of Feynman graphs is avoided.
For example, at tree-level, we need to compute only one Feynman diagram and we obtain \footnote{We redefine the coupling constant
$g =  g_{\rm YM} / \sqrt{2}$ in order to compare to results in the conventions of \cite{Anastasiou:2003kj,Bern:2005iz}. Also, we omit writing the
momentum conservation delta function $\delta^{(4)}(p_{1} + p_{2} + p_{3} +p_{4})$.}
\begin{equation}
A_{4}^{\rm tree} =  i g_{\rm YM}^2\,.
\end{equation}
The corresponding one-loop calculation is carried out in appendix \ref{appendix-oneloop}.
Introducing the notation
\begin{equation}
A_{4} = A_{4}^{\rm tree}  M_{4}\,,
\end{equation}
and using the result (\ref{app-oneloopfeynman}) we obtain
\begin{equation}\label{oneloop4pt}
M_{4} =  1 - \frac{a}{2}  {I}^{(1)}(s,t,m_{i}) + O(a^2)\,,
\end{equation}
where $s=(p_{1}+p_{2})^2$, $t=(p_{2}+p_{3})^2$ are the usual Mandelstam variables, $m_{i}$ are the Higgs masses introduced
in the previous section, and $a=g_{\rm YM}^2 N/(8 \pi^2)$,  with $g_{\rm YM}$ being the Yang-Mills coupling constant.\\

\begin{figure}
\psfrag{p1}[cc][cc]{$p_{2}$}
\psfrag{p2}[cc][cc]{$p_{3}$}
\psfrag{p3}[cc][cc]{$p_{4}$}
\psfrag{p4}[cc][cc]{$p_{1}$}
\psfrag{i1}[cc][cc]{\tiny $i_{2}$}
\psfrag{i2}[cc][cc]{\tiny $i_{3}$}
\psfrag{i3}[cc][cc]{\tiny$i_{4}$}
\psfrag{i4}[cc][cc]{\tiny$i_{1}$}
\psfrag{n}[cc][cc]{\tiny$j$}
\psfrag{x1m1}[cc][cc]{\tiny$(x_{2},m_{2})$}
\psfrag{x2m2}[cc][cc]{\tiny$(x_{3},m_{3})$}
\psfrag{x3m3}[cc][cc]{\tiny$(x_{4},m_{4})$}
\psfrag{x4m4}[cc][cc]{\tiny$(x_{1},m_{1})$}
\psfrag{x5m5}[cc][cc]{\tiny$(x_{5},0)$}
\psfrag{A}[cc][cc]{(a)}
\psfrag{B}[cc][cc]{(b)}
\psfrag{void}[cc][cc]{\phantom{text}}
 \centerline{{\epsfysize6cm
\epsfbox{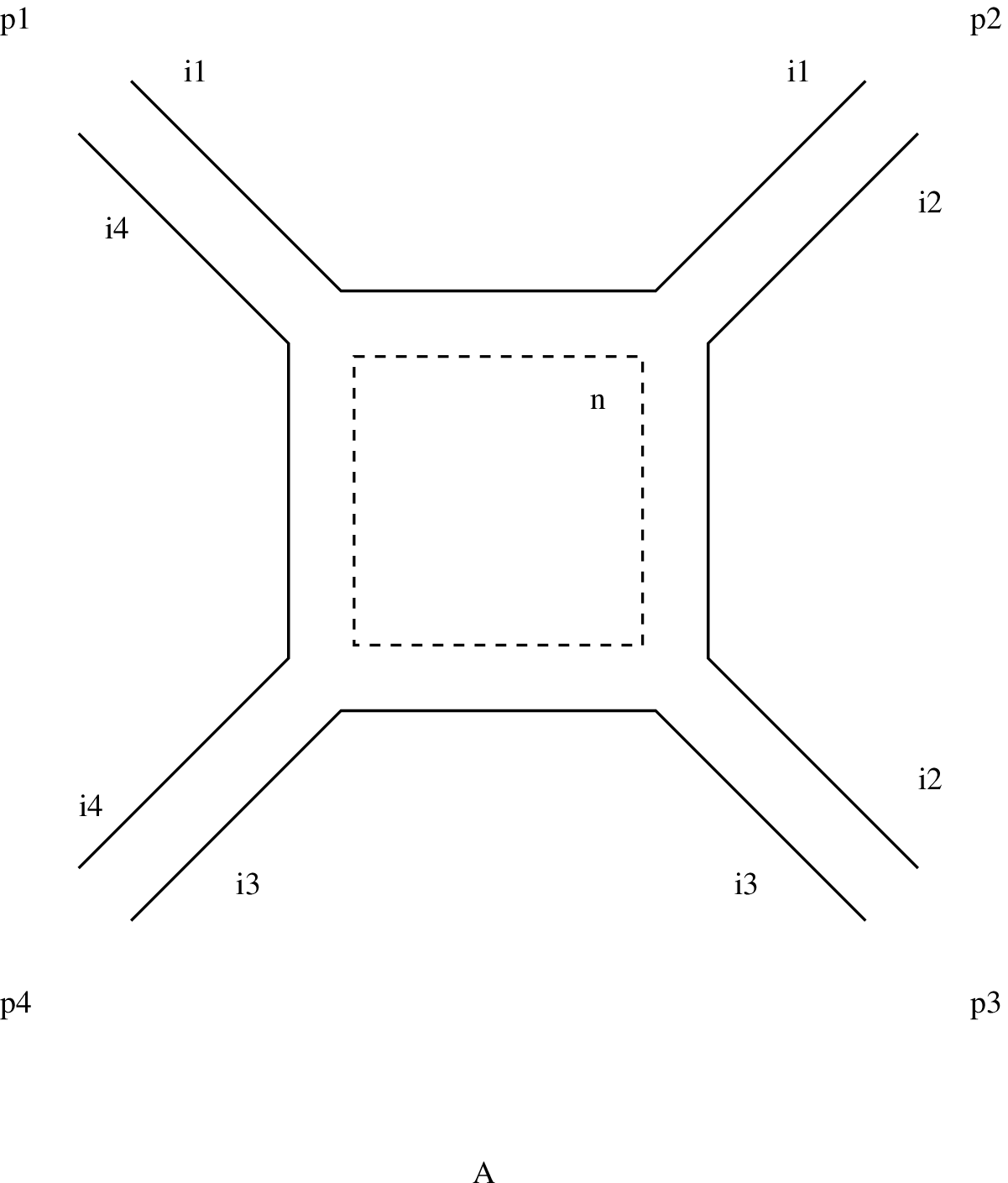}}
\hspace{2cm}
{\epsfysize6cm
\epsfbox{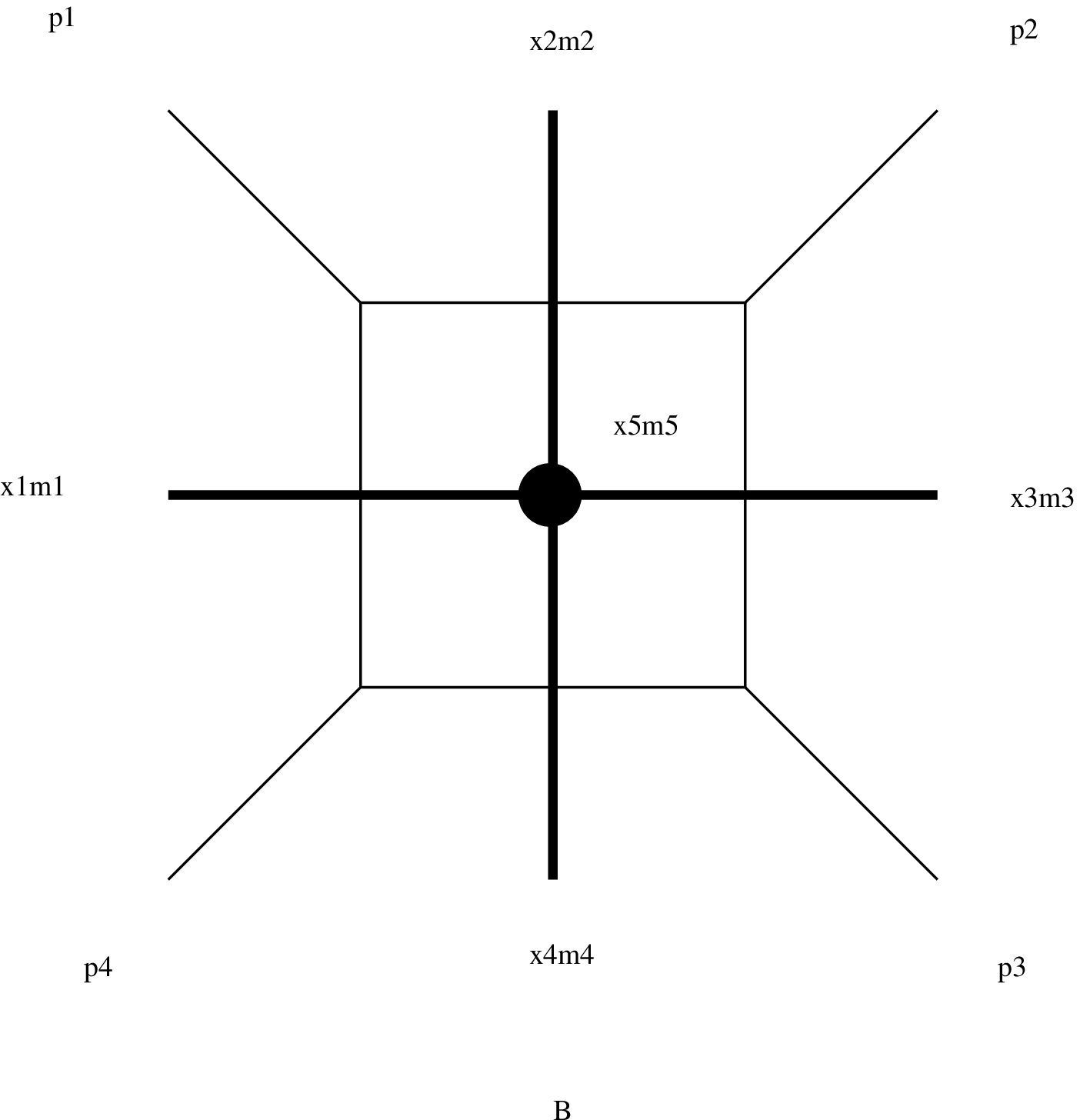}}
} \caption{\small
(a) Double line notation of the gauge factor corresponding to a one-loop box integral.
The $U(M)$ indices $i_{n}$ determine the masses of the different propagators.
(b) Dual diagram (thick black lines) and dual coordinates. The fifth component of the dual coordinates
corresponds to the radial AdS${}_{5}$ direction.}
\label{ftwo}
\end{figure}

The integral ${I}^{(1)}$ is a box integral, depicted in figure \ref{ftwo}. In contrast to dimensional
regularisation, it is defined in four dimensions and depends on several masses
coming from the Higgs mechanism.
The integral is given by
\begin{equation} \label{int0}
{I}^{(1)}(s,t,m_{i}) = c_{0} \, \int d^{4}k \frac{ ( s + (m_{1}-m_{3})^2)  (t + (m_{2}-m_{4})^2)}{(k^2+m_{1}^2) ( (k+p_{1})^2+m_{2}^{2}) ((k+p_{1}+p_{2})^2+m_{3}^2)(k-p_{4})^2+m_{4}^2)}\,.
\end{equation}
Here
$c_{0} =-i/\pi^2$.
{}From section \ref{sect-higgs} and appendix \ref{sect-propagators}, we see that the external masses are
\begin{equation}\label{extmasses}
p_{i}^2 = -(m_{i}-m_{i+1})^2\,.
\end{equation}
As was explained in section \ref{sect-strings}, in the string theory picture
the $m_{i}$ correspond to the distances between
the branes in the stack of $M$ branes and the $N$ branes. The scattering amplitude (say, of $M$ gluons)
corresponds to strings stretched between the different $M$ branes, with $i$ numbering the consecutive
gluons. Since two branes $i$ and $i+1$ are separated by $m_{i}-m_{i+1}$, the string connecting them
should have mass $|m_{i}-m_{i+1}|$ (in appropriate string units)  \cite{Kawai:2007eg}.
This situation corresponds precisely to the breaking of $U(N+M)$ to $U(N) \times U(1)^{(M-1)}$.
Later, we will take all masses equal, and the external momenta will become light-like in this limit (and we restore $U(N) \times U(M)$).
However keeping the masses distinct will allow us to make an interesting observation, as we will see presently.\\

In \cite{Drummond:2006rz} it was observed that the analogue of the above box integral in dimensional regularisation
has a broken dual conformal symmetry. This symmetry was made manifest by introducing
dual coordinates whose differences are the momenta of the scattered particles,
\begin{align}
k &= x_{5}-x_{1} =: x_{51}\,,& p_{1}&=x_{12} \,,& p_{2}&=x_{23} \,,& p_{3}&=x_{34}\, ,
\quad p_{4}=x_{41}\,.
\end{align}
Carrying out this change of variables in ${I}^{(1)}$ we obtain
\begin{equation}
 {I}^{(1)}(s,t,m_{i}) = c_{0} \,  \int d^{4}x_{5} \frac{(x_{13}^2 + (m_{1}-m_{3})^2) (x_{24}^2 + (m_{2}-m_{4})^2)}{(x_{15}^2+m_{1}^2)(x_{25}^2+m_{2}^2)(x_{35}^2+m_{3}^2)(x_{45}^2+m_{4}^2)}\,.
\end{equation}
{}From the discussion in section \ref{sect-strings}, it is natural to think of the masses as the
fifth components of the coordinates in the T-dual AdS${}_{5}$ space.
Therefore, let us define five-dimensional vectors $\hat{x}^{M}$, with $M=0\ldots 4$, and denote
the usual four-dimensional vectors by $x^{\mu}$, with $\mu=0 \ldots 3$.
Then, we define
\begin{align}
\hat{x}_{i}^{\mu} &:= x_{i}^{\mu}\,,& \hat{x}_{i}^{4} &:= m_{i}\,,& i&=1\ldots 4\,,
\end{align}
which allows us to rewrite ${I}^{(1)}$ as
\begin{equation}\label{int-fivedim}
 {I}^{(1)}(s,t,m_{i}) =  c_{0} \, \hat{x}^2_{13} \hat{x}^2_{24} \int d^{5}\hat{x}_{5} \frac{\delta(\hat{x}_{5}^{M=4})}{ \hat{x}_{15}^2 \hat{x}_{25}^2 \hat{x}_{35}^2 \hat{x}_{45}^2}\,.
\end{equation}
Here, the one-dimensional delta function was introduced for convenience. It enables us to write
the denominator of the integral in terms of five-dimensional, `hatted', quantities only.
Notice that, importantly, due to (\ref{extmasses}) we have that
\begin{equation}\label{five-dim-lightlike}
\hat{x}_{12}^2 = \hat{x}_{23}^2 = \hat{x}_{34}^2 = \hat{x}_{41}^2 =0\,.
\end{equation}
We note that these conditions are invariant under inversions in the five-dimensional space,
\begin{equation}\label{inv-fivedim}
\hat{x}_{i} \rightarrow \frac{\hat{x}_{i}}{\hat{x}_{i}^2}\,.
\end{equation}
Note that (\ref{inv-fivedim}) implies that $m_{i} \rightarrow m_{i}/\hat{x}_{i}^2$.
Moreover, in the form (\ref{int-fivedim}) it is also obvious that ${I}^{(1)}$ is invariant under the inversions (\ref{inv-fivedim}).
Indeed, in order to see the invariance of the integral in (\ref{int-fivedim}) it suffices to count the conformal weight of the various terms in  (\ref{int-fivedim}).
Importantly, as for the integrals discussed in \cite{Drummond:2006rz}, the conformal weight of the integration point is zero.
Moreover, the integral is normalised in such a way that the conformal weight at the external points is also zero, and
hence the integral is invariant under (\ref{inv-fivedim}).
By the same reasoning, one can see that e.g. triangle integrals would not be invariant.
Indeed, in the calculation leading to (\ref{oneloop4pt}),
all triangle integrals cancelled out.
This confirms that the symmetry expected from the string theory argument is present in the four-point amplitude we computed.\\

In addition to invariance under inversions, we have invariance under dilatations, and {\it four-dimensional} translation and rotation symmetry \footnote{Note that
because we have four-dimensional Lorentz symmetry only it would be mistaken to conclude that
the only allowed conformal invariants are five-dimensional cross-ratios.}.
The statement of the invariance of the integral under dual conformal transformations can of
course also be written in terms of differential equations.
The infinitesimal form of the dual conformal transformations is given in the appendix \ref{infconf}.
In particular we have
\begin{equation}\label{exact-dualconformal1loop}
\hat{K}_{\mu}  {I}^{(1)}(s,t,m_{i})  := \sum_{i=1}^{4}  \left[  2 x_{i}{}_{\mu} \left( x_{i}^{\nu} \frac{\partial}{\partial x_{i}^{\nu}} + m_{i} \frac{\partial}{\partial m_{i}} \right) - (x_{i}^2 +m_{i}^2) \frac{\partial}{\partial x_{i}^{\mu}} \right]  {I}^{(1)}(s,t,m_{i})  =0\,.
\end{equation}
Let us stress that this is an {\it exact} symmetry, i.e. there is no anomaly term on the
r.h.s. of (\ref{exact-dualconformal1loop}).
{\bf Three} remarks are in order here. Firstly, imagine expanding an arbitrary function $f(s,t,m_{i}=m \, \alpha_{i})$ for small $m$,
and truncating this expansion at some order.
Then, looking at the explicit form of $\hat{K}_{\mu}$, we see
that if $\hat{K}_{\mu}  f =0$ then the truncated expansion will have the same property, up to higher order terms in the expansion parameter.
Secondly, we remark that although the dual conformal symmetry is valid for genuine values of the Higgs masses $m_{i}$,
restricting the amplitude to the equal mass case $m_{i}=m$ will break the dual conformal symmetry.
Indeed, for dual conformal symmetry to work, it is important that $\hat{K}_{\mu}$ in $(\ref{exact-dualconformal1loop})$
can act on the different masses $m_{i}$. \footnote{Phrased differently, the equal mass configuration is not stable under dual conformal transformations,
as the latter would lead to a configuration with different masses.}
We will come back to this point in section \ref{label-Wardidentity} (see also the first reference in \cite{AWLreviews}).
Thirdly, in the conventional sense this exact dual conformal symmetry of the scattering amplitude
is not really a symmetry, as it acts on the masses $m_{i}$ and hence maps ${\cal N}=4$ super Yang-Mills theories at different points of moduli space into each other. While unconventional
from the field theory point of view, this mapping is nothing but an 
isometry in the dual string theory,
where the masses $m_{i}$ are coordinates in the fifth dimension of the dual space, as was
discussed in section two.
\\

{}From the string theory argument given in section \ref{sect-strings}, we expect this dual conformal symmetry to be a
generic property of scattering amplitudes in the Higgsed version of $\mathcal{N}=4$ SYM,
independently of the coupling constant and the number of external legs (see also sections \ref{label-higherloops} - \ref{sect-morelegs}).\\

This symmetry immediately allows us to make the observations of \cite{Drummond:2006rz}
more precise and useful. As was already discussed, triangle (sub-)diagrams are excluded
and only the restricted set of dual conformal integrals (with the mass assignments as explained in
section \ref{sect-higgs}) are allowed in the final answer for a scattering amplitude.
In practice, once one has identified those integrals for the scattering amplitude under
consideration, e.g.~(\ref{int-fivedim}) in the one-loop and
(\ref{doublebox}) in the two-loop case, one can set $m_{i}=m$ in order to simplify
the calculation of the integrals.
Moreover, one can consider the small $m$ expansion and neglect any terms evanescent in $m^2$.\\

We write down invariants of this five-dimensional dual conformal symmetry.
In the generic $n$-point case, we can start from
the four-dimensional Lorentz invariants $x_{ij}^2$. It is easy to see that they can be turned
into dual conformal invariants by defining
\begin{equation}
u_{ij} := \frac{ m_{i} m_{j} } {\hat{x}_{ij}^2 }\,.
\end{equation}
Note that $u_{i,i+1}$ is ill defined in view of the light-likeness conditions (\ref{five-dim-lightlike}).
Therefore, we can have the following two conformal invariants in the four-point case,
\begin{align}
u&:= u_{13} =\frac{ m_{1} m_{3}}{\hat{x}_{13}^2}\,,& v:=u_{24}= \frac{m_{2} m_{4}}{\hat{x}_{24}^2}\,.
\end{align}
Hence we arrive at the non-trivial statement that
\begin{equation}\label{conf-answer}
 {I}^{(1)}(x_{13}^2,x_{24}^2,m_{i}) = f\left(\frac{ m_{1} m_{3}}{\hat{x}_{13}^2},\frac{m_{2} m_{4}}{\hat{x}_{24}^2}\right)\,.
\end{equation}
As a consequence of dual conformal symmetry, the integral with four different masses is
reduced to a two-variable function.
The relevant four-point integral is given in \cite{'tHooft:1978xw}, and it can be checked that the known answer for ${I}^{(1)}$ is in agreement with (\ref{conf-answer}).\\

If we think about the masses $m_{i}$ as regulating the amplitude, then it is interesting to know the integral ${I}^{(1)}$  for the equal mass
case $m_{i}=m$ and $m$ small compared to the kinematical variables $s$ and $t$.
If we did not know the result of \cite{'tHooft:1978xw}, we could carry out a simpler calculation for $m_{i}=m$ and obtain
\begin{equation}\label{box-param}
 {I}^{(1)}(x_{13}^2,x_{24}^2,m)=  2 \ln \left( \frac{ m^2}{x_{13}^2} \right) \, \ln\left( \frac{m^2}{x_{24}^2} \right) - \pi^2+ O(m^2) \,.
\end{equation}
We remark that from (\ref{box-param}) it follows that the function $f$ in (\ref{conf-answer}) is given by
$
f(u,v) =
2 \ln(u)  \ln(v) - \pi^2 + O(m^2)
$.

\subsection{Higher loops and four-point exponentiation}
\label{label-higherloops}
\begin{figure}
\label{fthree}
\psfrag{p1}[cc][cc]{$p_{2}$}
\psfrag{p2}[cc][cc]{$p_{3}$}
\psfrag{p3}[cc][cc]{$p_{4}$}
\psfrag{p4}[cc][cc]{$p_{1}$}
\psfrag{i1}[cc][cc]{\tiny $i_{2}$}
\psfrag{i2}[cc][cc]{\tiny $i_{3}$}
\psfrag{i3}[cc][cc]{\tiny$i_{4}$}
\psfrag{i4}[cc][cc]{\tiny$i_{1}$}
\psfrag{n}[cc][cc]{\tiny$j$}
\psfrag{m}[cc][cc]{\tiny$k$}
\psfrag{x1m1}[cc][cc]{\tiny$(x_{2},m_{2})$}
\psfrag{x2m2}[cc][cc]{\tiny$(x_{3},m_{3})$}
\psfrag{x3m3}[cc][cc]{\tiny$(x_{4},m_{4})$}
\psfrag{x4m4}[cc][cc]{\tiny$(x_{1},m_{1})$}
\psfrag{x5m5}[cc][cc]{\tiny$(x_{5},0)$}
\psfrag{box2aA}[cc][cc]{(a)}
\psfrag{labelbox2m}[cc][cc]{(b)}
 \centerline{{\epsfysize5cm
\epsfbox{box2a.eps}}
\hspace{2cm}
{\epsfysize5cm
\epsfbox{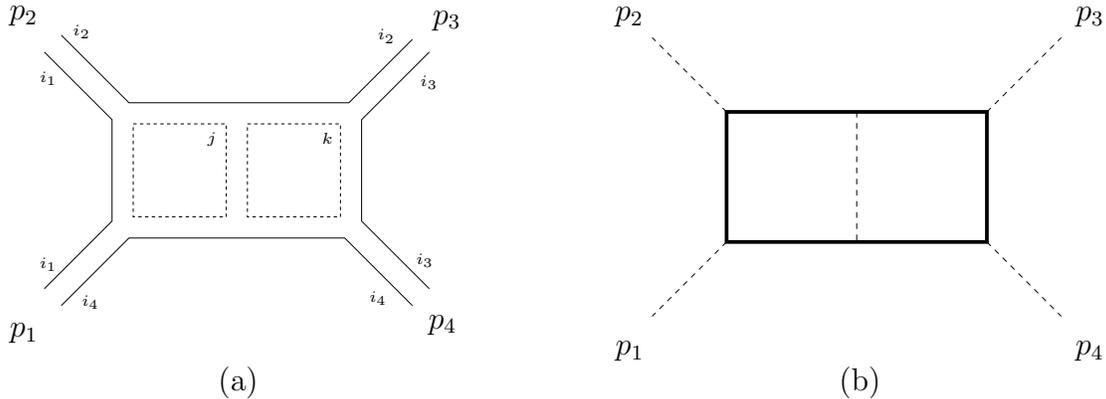}}
\vspace{0.5cm}
}
\caption{\small
(a) Double line notation of the gauge factor corresponding to the two-loop box integral in the Higgsed theory. The integral is
dual conformally invariant.
(b) Diagram for the same integral in the equal mass case $m_{i}=m$. Dashed thin lines denote
massless propagators, thick black lines denote massive propagators.}
\end{figure}

If the inversion symmetry found in section \ref{S-one-loop} is present at any loop order
then it dramatically restricts the set of scalar integrals that can appear. We would basically find
the integrals considered in \cite{Drummond:2006rz}, with the difference that the outer
loop carries masses, with the mass assignments as explained in
section \ref{sect-higgs}. E.g.~at two loops we expect to find the following integral only (cf. figure \ref{fthree}),
\begin{equation}\label{doublebox}
{I}^{(2)}(s,t,m_{i}) = (c_{0})^2 \,   (\hat{x}_{13}^2)^2  (\hat{x}_{24}^2) \int d^{5}\hat{x}_{5}  \int d^{5}\hat{x}_{6}  \frac{\delta(\hat{x}_{5}^{M=4}) \delta(\hat{x}_{6}^{M=4})  }{ \hat{x}_{15}^2 \hat{x}_{25}^2 \hat{x}_{35}^2 \hat{x}_{56}^2 \hat{x}_{36}^2 \hat{x}_{46}^2   \hat{x}_{16}^2}\,,
\end{equation}
where $\hat{x}_{i,i+1}^2 = 0$ as in the one-loop case.
The momentum space notation may be more familiar to some readers, which in the
equal mass case is given by
\begin{align}\label{doubleboxequalmass}
{I}^{(2)}(s,t,m) &= (c_{0})^2 \, s^2 t \int d^{4}k_{1} \int  d^{4}k_{2}  \big[ P(k_{1},m^2) P(k_{1}+p_{1},m^2) P(k_{1}+p_{1}+p_{2},m^2)  \nonumber \\
 &\=\phantom{(c_{0})^2 \,  } \times P(k_{1}-k_{2},0) P(k_{2},m^2) P(k_{2}-p_{4},m^2) P(k_{2}-p_{3}-p_{4},m^2) \big] \,,
\end{align}
where $P(k,m^2) = (k^2 + m^2)^{-1}$ and the external momenta are light-like, $p_{i}^2=0$.
The double box integral may also appear in a different orientation obtained by replacing $\hat{x}_{1} \to \hat{x}_{2}\,, \ldots\,, \hat{x}_{4} \to \hat{x}_{1}$,
which amounts to interchanging $s$ and $t$ in (\ref{doubleboxequalmass}).
We argue that the coefficients of the box integrals must be the same as those obtained in dimensional
regularisation \cite{Anastasiou:2003kj,Bern:2005iz}. The reason is that the leading infrared divergence cannot depend on the regularisation.
Therefore, based on dual conformal symmetry we expect \footnote{For convenience, we write the following formulae in the equal mass case $m_{i}=m$. Note
that one can always restore the full dependence on the $m_{i}$ by substituting $m^2/s  \to m_{1} m_{3} / \hat{x}_{13}^2$ and similarly  $m^2/t \to m_{2} m_{4} / \hat{x}_{24}^2$,
thanks to dual conformal symmetry.}
\begin{equation}
M_{4} = 1 - \frac{a}{2} {I}^{(1)}(s,t,m) + \frac{a^2}{4}  \left[ {I}^{(2)}(s,t,m) +  {I}^{(2)}(t,s,m) \right] + O(a^3) \,,
\end{equation}
with $a=g_{\rm YM}^2 N /(8 \pi^2)$.
Following  \cite{Anastasiou:2003kj,Bern:2005iz}, we compute \begin{equation}
\ln M_{4}  = a \, w^{(1)} + a^2  \, w^{(2)} + O(a^3)\,,
\end{equation} in order to see
whether we find exponentiation in our Higgs regularisation.
It is convenient to write all quantities that appear in a small $m^2$ expansion in the following form,
\begin{equation}
f(s,t,m^2) = \sum_{i=1}^{i_{\rm max}} \left[ \ln^i(m^2/s) + \ln^i(m^2/t) \right] \,  f_{i}(s/t) + f_{0}(s/t) + O(m^2)\,.
\end{equation}
At one loop, we find, using (\ref{box-param}),
\begin{equation}\label{woneloop}
w^{(1)} = -\frac{1}{2} \left[ \ln^2(m^2/s) + \ln^2(m^2/t) \right] +\frac{1}{2} \ln^2(s/t) +\frac{1}{2} \pi^2  + O(m^2) \,.
\end{equation}
At two loops, using equations 
(\ref{appendix-I2stsymm}) and
(\ref{appendix-I1squared}) of appendix  C we obtain
\begin{align}\label{wtwoloops}
w^{(2)} &= -\frac{1}{8} (I^{(1)}(s,t,m))^2  + \frac{1}{4} {I}^{(2)}(s,t,m) + \frac{1}{4} {I}^{(2)}(t,s,m)   \\
&=  \frac{1}{2} \zeta_{2} \left[ \ln^2(m^2/s) + \ln^2(m^2/t) \right]   - \zeta_{3}  \left[ \ln(m^2/s) + \ln(m^2/t) \right] + \left[  - \frac{1}{2} \zeta_{2} \ln^2(s/t)-\frac{3}{40} \pi^4  \right] + O(m^2)  \,, \nonumber
\end{align}
where various terms cancelled when taking the logarithm.
Let us now discuss these results.\\

To begin with, in analogy with dimensional regularisation, we define the cusp anomalous dimension by (see \cite{Drummond:2007aua}
and references therein)
\begin{equation}\label{defgammacusp}
\left( \frac{\partial}{\partial \ln(m^2)} \right)^2  \ln M_{4} = - \Gamma_{\rm cusp}(a) + O(m^2)\,.
\end{equation}
Plugging in the explicit results (\ref{woneloop}) and (\ref{wtwoloops}) into (\ref{defgammacusp})
we find
\begin{equation}
\Gamma_{\rm cusp}(a) = 2 a - 2 \zeta_{2}  a^2 + O(a^3)\,,
\end{equation}
in agreement with the expression in dimensional regularisation.
{
Next, we check whether the finite part of the two-loop result can be thought of as the exponentiation of the finite part of the one-loop result,
as in dimensional regularisation.
 Indeed, let us define a finite part $F_{4}$ of $\ln M_{4}$ according to 
 \begin{equation}\label{lnM4}
 \ln M_{4} = D_{4} + F_{4} +O(m^2) \,.
 \end{equation}
 Here $D_{4}$ contains the terms associated to the infrared divergences,
\begin{equation}\label{defD4}
D_{4} = - \frac{1}{4} \Gamma_{\rm cusp}(a) \, \left[ \ln^2(m^2/s) + \ln^2(m^2/t) \right]   + G(a) \,  \left[ \ln(m^2/s) + \ln(m^2/t) \right] \,,
\end{equation} 
where we have introduced the `collinear anomalous dimension' $G(a) = - \zeta_{3} a^2 + O(a^3)$.
Note that $F_{4}$ is a function of $s/t$ (and of the coupling $a$) and that it is defined up to an additive (coupling-dependent) constant.
The value of this constant and that of $G(a)$ in (\ref{defD4}) 
are scheme dependent and can be modified by a redefinition
\begin{equation}\label{scheme-dep}
m^2 \to m^2 \,  e^{h(a)}\,,
\end{equation}
where $h(a)$ is an arbitrary function. \\

Let us now expand equation (\ref{lnM4}) in the coupling constant. Writing
 $F_{4} = a \, F_{4}^{(1)} + a^2 \, F_{4}^{(2)} + O(a^3)$, we obtain (cf. (\ref{woneloop}) and (\ref{wtwoloops}))
\begin{align}
F_{4}^{(1)} &=  \frac{1}{2} \ln^2(s/t) +\frac{1}{2} \pi^2 \,,& F_{4}^{(2)} &= - \frac{1}{2} \zeta_{2} \ln^2(s/t)-\frac{3}{40} \pi^4  \,.
\end{align}
Combining these results we see that, up to two loops,
\begin{equation}
F_{4} = \frac{1}{2} \Gamma_{\rm cusp}(a)  F_{4}^{(1)} + C(a)  \,,
\end{equation}
just as in dimensional regularisation, and in agreement with the
anomalous dual conformal Ward identity derived in \cite{Drummond:2007au,Drummond:2008vq}.
We will comment on the relation between the exact dual conformal symmetry in the Higgs regularisation
and that anomalous Ward identity in section \ref{label-Wardidentity}.
We find that $C(a)=  \frac{1}{120}\pi^4 \, a^2 + O(a^3)$.
}\\

To summarise, we see that taking only the integrals allowed by dual conformal symmetry
at two loops agrees with all features discovered for the corresponding amplitudes
computed in dimensional regularisation.\\

One can extend the analysis presented here to higher loops and more external
legs. Interestingly, it was technically quite simple to evaluate the small mass
expansion of the double box integral (see appendix \ref{appendix-integrals}).
It is desirable to automatise the method used to compute that integral
and to apply it to more complicated cases.

\subsection{Anomalous dual conformal Ward identity vs. exact dual conformal symmetry}
\label{label-Wardidentity}
We argued that the scalar four-point scattering amplitudes in the Higgsed version of $\mathcal{N}=4$ SYM
\eqn{scalar4ptamplitude} should have an exact dual conformal symmetry,
i.e. they should satisfy the equation
\begin{equation}\label{exact-dualconformal}
\hat{K}_{\mu} M_{4} := \sum_{i=1}^{4}  \left[  2 x_{i}{}_{\mu} \left( x_{i}^{\nu} \frac{\partial}{\partial x_{i}^{\nu}} + m_{i} \frac{\partial}{\partial m_{i}} \right) - (x_{i}^2 +m_{i}^2) \frac{\partial}{\partial x_{i}^{\mu}} \right] M_{4} =0\,,
\end{equation}
where we recall that $A_{4} = A_{4}^{\rm tree} M_{4}$.
Notice that an equation very similar to (\ref{exact-dualconformal}) has already appeared at strong coupling in the first reference of \cite{AWLreviews}.
It seems natural to ask what the relation between (\ref{exact-dualconformal}) and the anomalous dual conformal Ward identity
of  \cite{Drummond:2007au,Drummond:2008vq} is, namely
\begin{equation}\label{broken-dualconformal}
{K}_{\mu} F_{n}  :=\sum_{i=1}^{n}  \left[  2 x_{i}{}_{\mu}  x_{i}^{\nu} \frac{\partial}{\partial x_{i}^{\nu}}  - x_{i}^2 \frac{\partial}{\partial x_{i}^{\mu}} \right] F_{n} = \frac{1}{2} \Gamma_{\rm cusp}(a) \sum_{i=1}^{n}  \left[ x_{i,i+1}^{\mu} \, \ln \frac{x^2_{i,i+2}}{x^2_{i-1,i+1}} \right] F_{n}\,,
\end{equation}
where $F_{n}$ is defined as the finite part (i.e., with the logarithm, i.e. $\ln^2 m^2 , \ln m^2 $, terms removed) of $\ln M_{n} $.
Equation (\ref{broken-dualconformal}) was initially derived in \cite{Drummond:2007au} for certain light-like Wilson loops dual to maximally-helicity-violating amplitudes
by analysing the structure of divergences of the latter, which leads to the appearance of the anomalous term on the r.h.s. of (\ref{broken-dualconformal}). \\

Now, notice that in equation (\ref{exact-dualconformal}) we could replace $M_{4}$ by $\ln M_{4}$. Then, splitting up $\ln M_{4} = D_{4} + F_{4} + O(m_{i}^2)$
into a divergent and a finite part, it is clear that the action of the differential operator on the l.h.s. of (\ref{exact-dualconformal}) on $D_{4}$
will produce an anomalous term \footnote{Note that acting on $F_{n}$, which by definition is independent of the regulator $m_{i}$, we
have $\hat{K}_{\mu} F_{n} = K_{\mu} F_{n} + O(m_{i}^2)$, and we can simply replace $\hat{K}_{\mu}$ by $K_{\mu}$.}.
Although this is not obvious, we expect the quantity 
 $F_{4}$ to be independent of the regularisation method that was used to calculate it (up to a scheme-dependent additive constant, as was discussed
 in the previous section).
Therefore, we expect that $F_{4}$ computed in the Higgsed theory should satisfy the same anomalous Ward identity (\ref{broken-dualconformal}) as when computed
in dimensional regularisation. We indeed see that this is the case in the two-loop example considered in section \ref{label-higherloops}, as one
can easily check.
There is little doubt that  one can prove that (\ref{broken-dualconformal}) follows from (\ref{exact-dualconformal}) by studying
the structure of divergences of scattering amplitudes in the Higgsed theory (with different masses).

\subsection{More external legs}
\label{sect-morelegs}

\begin{figure}
\psfrag{p1}[cc][cc]{$p_{2}$}
\psfrag{p2}[cc][cc]{$p_{3}$}
\psfrag{p3}[cc][cc]{$p_{4}$}
\psfrag{p4}[cc][cc]{$p_{5}$}
\psfrag{p5}[cc][cc]{$p_{1}$}
\psfrag{i1}[cc][cc]{\tiny $i_{2}$}
\psfrag{i2}[cc][cc]{\tiny $i_{3}$}
\psfrag{i3}[cc][cc]{\tiny$i_{4}$}
\psfrag{i4}[cc][cc]{\tiny$i_{5}$}
\psfrag{i5}[cc][cc]{\tiny$i_{1}$}
\psfrag{n}[cc][cc]{\tiny$j$}
\psfrag{n2}[cc][cc]{\tiny$k$}
\psfrag{x1m1}[cc][cc]{\tiny$(x_{2},m_{2})$}
\psfrag{x2m2}[cc][cc]{\tiny$(x_{3},m_{3})$}
\psfrag{x3m3}[cc][cc]{\tiny$(x_{4},m_{4})$}
\psfrag{x4m4}[cc][cc]{\tiny$(x_{1},m_{1})$}
\psfrag{x5m5}[cc][cc]{\tiny$(x_{5},0)$}
\psfrag{labell1b1}[cc][cc]{(b)}
\psfrag{labell1b1colour}[cc][cc]{(a)}
 \centerline{{\epsfysize6cm
\epsfbox{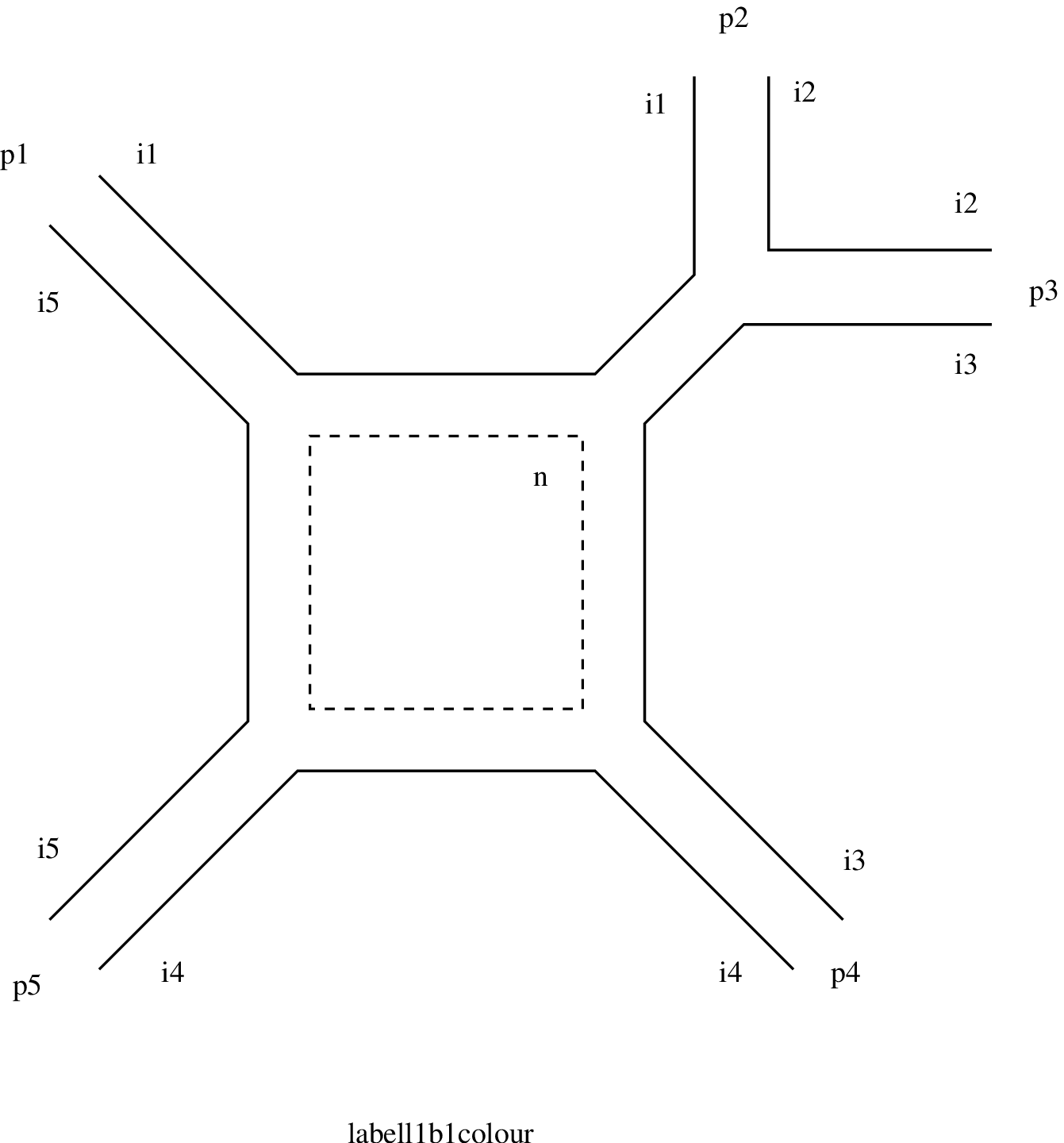}}
\hspace{2cm}
{\epsfysize6cm
\epsfbox{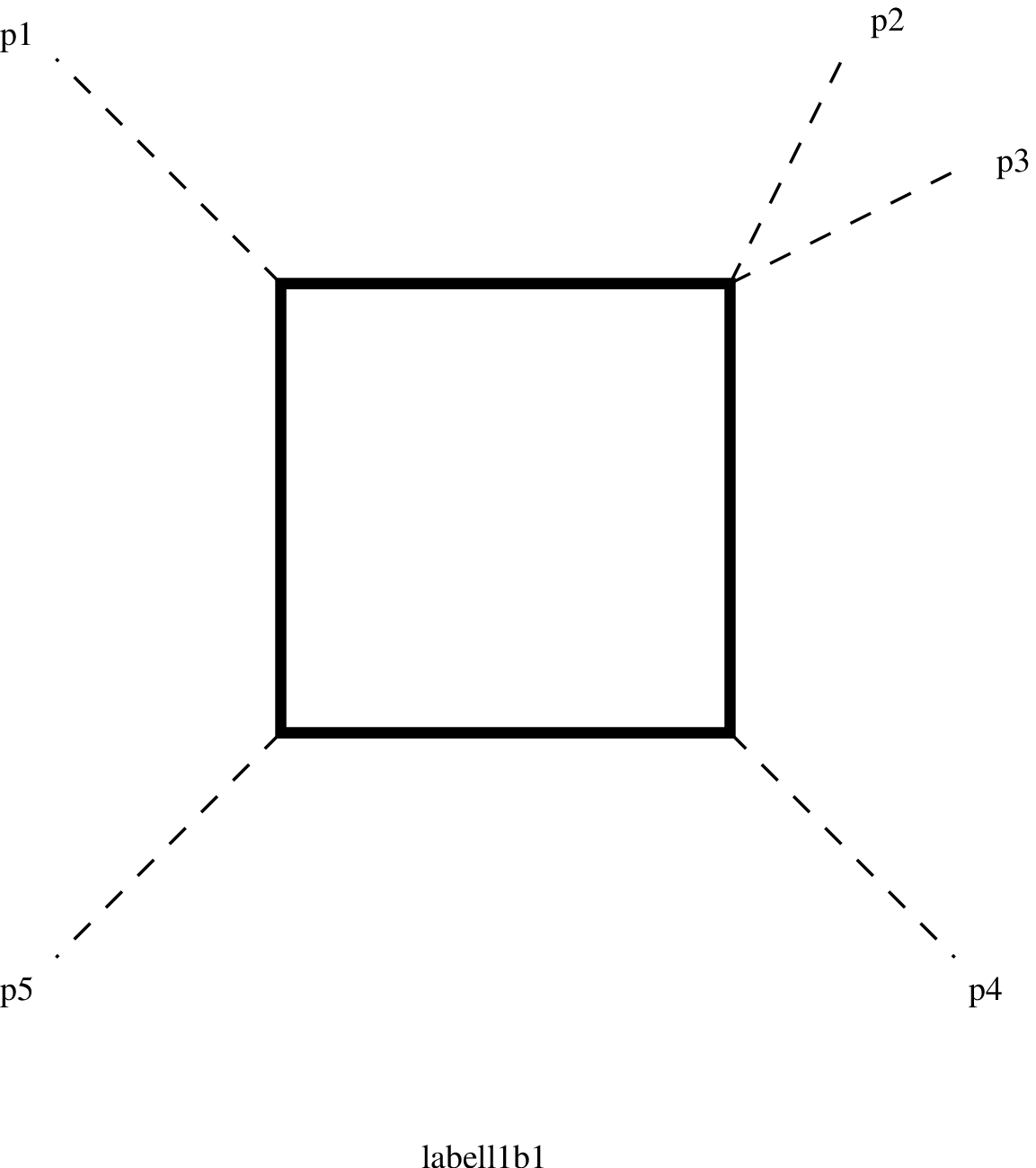}}
\vspace{0.5cm}
}  \caption{ 
\small
(a) An example of a higher-point dual conformal integral. The picture corresponds to a `1-mass' integral, since the sum $p^{\mu}_{3}+p^{\mu}_{4}$ is in general
not light-like. As in the four-point case, there are the masses of the Higgsed particles circulating in the outer loops.
(b) In the equal mass case $m_{i}=m$, all outer legs become massless (dashed lines), while the internal propagators (full black lines) have uniform mass $m$,
making the integral infrared finite.}
\label{figure-fivepoints}
\end{figure}

Turning to the generic $n$-point case,
we would like to argue that at one loop the only effect of the new regularisation
is to replace the dimensionally regulated box integrals appearing in dimensional
regularisation by our mass regulated box integrals, with
the specific mass assignment explained earlier \footnote{When scattering particles with helicity, there will also be a slight change in the spinor helicity formalism
since in the distinct mass case the external states are massive with masses squared $(m_{i}-m_{i+1})^2$. We could argue that
this effect is irrelevant since we could consider a situation where  $(m_{i}-m_{i+1})^2 \ll m_{j}^2$.}.
See figure \ref{figure-fivepoints}, which illustrates the only five-point dual conformal scalar integral at one loop. The generalisation to an arbitrary
number of external legs is straightforward (see also appendix \ref{appendix-integrals}).\\

For this it is desirable to have a suitable $n$-leg generalisation of the four scalar amplitude
\eqn{scalar4ptamplitude} considered above at hand, which has the virtue of coming from
a single planar tree-diagram. We propose the non-MHV amplitude of  $2n$ external scalar fields,
\begin{equation}\label{2nexample2}
A_{2n}=\langle\Phi_{4}\,\underbrace{\Phi_{5}\,\Phi_{6}\,\Phi_{7}\,\Phi_{5}\,
\Phi_{6}\,\Phi_{7}\,\ldots}_{n-1}\,\Phi_{4}\,\underbrace{\ldots\,\Phi_{7}
\, \Phi_{6} \, \Phi_{5}\,\Phi_{7} \, \Phi_{6} \, \Phi_{5}}_{n-1} \rangle \,.
\end{equation}
Note that $A_{2n}$ is an N${}^{k}$MHV amplitude where $k=n-2$, with $n \ge 2$. For $n=2$ it is equivalent to
the four-scalar amplitude considered in equation (\ref{scalar4ptamplitude}).
In (\ref{2nexample2}) we suppressed the dependence on the scattering momenta $p_{1}\,, \ldots p_{n}$, and
the flavour choice of the scalars was made in a way such that there is only one tree-level diagram
(compare figure 6),
which can be readily evaluated, \footnote{We remark that (\ref{2nexample2}) may be very interesting in its own right. It appears to be
a new example of an amplitude that is dual conformal on its own, without having to consider superamplitudes.
In this sense, (\ref{2nexample2}) is very similar to split helicity amplitudes. Moreover, it is given by a single term.
This may suggest using an alternative formulation of tree-level amplitudes
(as compared to the one given in \cite{Drummond:2008cr}), where (\ref{2nexample2}) plays the role of the starting point.
We thank J.~Drummond for discussions on this point.}
\begin{equation}\label{2ntree2}
A^{\rm tree}_{2n} = i g_{\rm YM}^{(2n-2)} \,\,  \frac{1} {\hat{x}_{2n,3}^2 \, \hat{x}_{2n-1,4}^2 \ldots \hat{x}_{n+3,n}^2 }  =i g_{\rm YM}^{(2n-2)}  \,\, \prod_{i=0}^{n-3} \frac{1}{\hat{x}^2_{2n-i,i+3}} \,.
\end{equation}
At one-loop level, the calculation of $A^{1-{\rm loop}}_{2n}$ would be very similar to that done for $A^{1-{\rm loop}}_{4}$ in appendix \ref{appendix-oneloop}.
 As a result, we expect that $A^{1-{\rm loop}}_{2n}$ will be given by
a linear combination of the dual conformal integrals given in appendix \ref{appendix-integrals}, with certain coefficients.

\begin{figure}
\psfrag{4}[cc][cc]{$4$}
\psfrag{5}[cc][cc]{$5$}
\psfrag{6}[cc][cc]{$6$}
\psfrag{7}[cc][cc]{$7$}
\centerline{{\epsfxsize7cm
\epsfbox{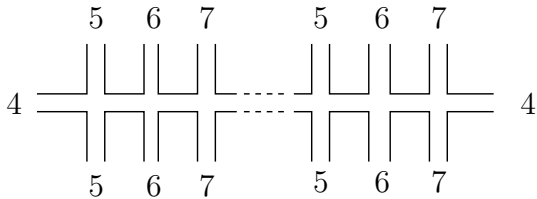}}
}  \caption{ 
\small  The unique planar tree-level diagram contributing to the family of multi-leg amplitudes $A_{2n}$ of (\ref{2nexample2}). Note that the flavour sequence $567\ldots567$ does not need to end on 
the $7$. }
\label{a2namp}
\end{figure}

\subsection{Dual conformal symmetry vs. dual superconformal symmetry}
\label{dualsuperconformal}
The four-point amplitude we considered in the previous sections
is special in the sense that it is very similar to the so-called
`split-helicity' case. This refers to the scattering of gluons where the
gluons with negative helicity sit on one side of the colour ordered amplitude and all gluons with
positive helicity sit on the other side. In \cite{Drummond:2008vq} it was shown
that these amplitudes are dual conformal on their own.
For generic helicity configurations, this is not true and one needs to consider
certain super-amplitudes which are dual conformal  \cite{Drummond:2008vq}. In particular, the dual
conformal generator then receives additional terms depending on Grassmann
variables that parametrise the on-shell states of the $\mathcal{N}=4$
on-shell supermultiplet.
Here we want to argue that the observation that
the integrals appearing in loop calculations have the exact conformal
symmetry discussed earlier applies to all amplitudes, not just to the split helicity case.
Indeed, the reason for the non-covariance of generic amplitudes under dual
conformal symmetry is that different amplitudes can transform into each other
under this symmetry. When one considers super-amplitudes as in  \cite{Drummond:2008vq},
the latter transform covariantly. However, knowing that a generic
amplitude can be expressed as a sum over scalar integrals multiplied by certain coefficients,
it is clear that the non-covariance under the dual conformal transformations can
affect the coefficients only. The integrals, on the other hand, should have the exact
dual conformal symmetry. A concrete analysis of this question, i.e. the supersymmetrization
of our construction including the 
necessary extension of the dual conformal generators with terms depending on suitably
defined Grassmann variables, is left for future work.

\section{Conclusion and outlook}
In this paper we investigated a regularisation of (planar) scattering amplitudes in $\mathcal{N}=4$ SYM
that is an alternative to the commonly used dimensional regularisation/reduction.
This regularisation was motivated by the string theory side of the AdS/CFT correspondence and, as for instance mentioned in \cite{Berkovits:2008ic} and argued in this paper, it also suggests
that the previously discovered broken dual conformal symmetry \cite{Drummond:2006rz,Alday:2007hr,Drummond:2007cf,Drummond:2007au} of
scattering amplitudes can be turned into an exact symmetry when considering scattering
amplitudes in the Higgsed theory.\\

We worked out the gauge theory analogue of this regularisation
and argued that the scattering amplitudes on the gauge theory side should possess the aforementioned exact dual conformal symmetry.
The latter severely restricts
the number and type of loop integrals that can appear in the calculation of the amplitudes. \footnote{In dimensional
regularisation, the observation that the integrals appearing in the four-gluon amplitude of \cite{Anastasiou:2003kj,Bern:2005iz} all have dual conformal properties
was made in \cite{Drummond:2006rz}. However, since the dimensional regularisation breaks this symmetry, other integrals could also appear in principle.
In contrast, in the case of the mass regularisation considered in this paper, the dual conformal symmetry is exact and hence so
is the restriction on the possible integrals appearing in the amplitude.}
In particular, dual conformal symmetry forbids all triangle sub-graphs. Furthermore, there are simple rules for
determining whether an integral is dual conformal, see \cite{Drummond:2006rz}.
Dual conformal symmetry is a very helpful tool for establishing the set of
scalar integrals that are allowed to appear in an amplitude.
Let us now comment on possible future directions.\\

We suggest that an analysis of recursion relations at tree level and generalised unitarity methods at loop
level in the Higgsed theory may be very interesting, at least for the following two reasons.
Firstly, it seems likely that one can prove the
exact dual conformal symmetry reported on in this paper recursively using the ideas presented in \cite{Brandhuber:2008pf,dualconf1loop}.
Since all integrals in our set-up are four-dimensional, it may be possible to prove the symmetry to arbitrary loop order.
Combined with a systematic understanding of the infrared divergences in the Higgsed theory this should lead to
a proof of the anomalous dual conformal Ward identity \footnote{The anomalous dual conformal Ward identity
was proven for Wilson loops in \cite{Drummond:2007cf,Drummond:2007au}, and it was conjectured in \cite{Drummond:2008vq} that it should hold
for arbitrary non-MHV amplitudes. Further evidence for this conjecture was collected in \cite{dualconf1loop}. }
 that appears when one separates the amplitude in a divergent
and a finite part.
Secondly, we argued that dual conformal symmetry is very helpful to find an integral basis for integrals that
are allowed to appear in a given amplitude. Once the integral basis has been determined,
the coefficients of the integrals have to be computed.
In higher-loop calculations, this is usually done by a method based on generalised unitarity  \cite{unitarity94} (for a review see \cite{OSrev}; see also \cite{Drummond:2008bq},
and also \cite{Bern:1995db} for an application with internal masses).
In the present case, all calculations can be done in exactly four dimensions. We are confident that this will
be an advantage when computing higher-loop and higher-leg amplitudes.\\

An important question we hope to address in the future is
whether the conventional conformal symmetry of $\mathcal{N}=4$ SYM
can be used to constrain scattering amplitudes at loop level.
It may be that it is easier to understand that symmetry in our regularisation.\\

It would also be interesting to study the Regge behaviour of scattering amplitudes in the Higgsed theory.
As was already stressed, in the Higgsed theory, taking the logarithm of an amplitude computed to a certain order in the coupling does not require knowing
the evanescent terms in the lower-loop amplitudes, contrary to dimensional regularisation.
It is natural to think that this feature would also make the analysis of Regge behaviour of scattering
amplitudes easier.
\\

A natural question is whether the scattering amplitudes in the Higgsed theory, at least in the
 maximally helicity violating case, can be reproduced by some kind of Wilson loop.
{}From the string theory point of view, it seems clear that such a Wilson loop description, if it exists, could only arise in the limit
 where the regulator is taken to be small.
On the gauge theory side, it appears fairly easy to engineer a Wilson loop that reproduces the one-loop $N$-point MHV amplitudes
in the equal mass case for small mass,
by inserting a  mass into the propagator as suggested in \cite{McGreevy:2007kt}. However, it is not clear whether a
Wilson loop could also reproduce the non-trivial $m_{i}$ dependence in the general case, where all
$m_{i}$ are small with respect to the momentum invariants.
Moreover, with the lack of a more physical motivation it is doubtful whether any of these `engineered'
agreements continue beyond one-loop order. It may be that the conjectured duality between MHV amplitudes
and Wilson loops holds in dimensional regularisation only, though one would expect such a duality (once clearly understood!) to be independent of the regularisation prescription.\\

Finally, as the suggestion for studying scattering amplitudes in the Higgsed theory came from the
AdS/CFT correspondence, one may wonder whether it is also useful to carry out computations on the
string theory side of the correspondence. On the string theory side, while this regularisation is conceptually very
appealing, it seems difficult to carry out actual computations. On the other hand, this
regularisation may be more amendable to systematically computing sub-leading corrections \footnote{See for instance \cite{Kruczenski:2007cy} regarding difficulties when using dimensional regularisation.} in $1/\sqrt{\lambda}$ and in order to answer questions related to the symmetries of the scattering amplitudes (where finding the classical solutions may not be necessary).

\section*{Acknowledgments}
We would like to thank L.~Dixon, J.~Drummond, H.~Elvang, T.~Klose, G.~Korchemsky,
J.~Maldacena,  S.~Moch, S.~Naculich and P.~Vieira for
discussions. J.H. thanks T.~Riemann and V.~Yundin for correspondence on \cite{Gluza:2007rt}, and
A.~Rodigast for computer support.
L.F.A. and J.H. thank the APCTP, Pohang, Korea, J.H. thanks the LAPTH, Annecy, France,
and J.P. thanks the PITP, University of British Columbia, Vancouver, Canada,
where part of this work was done, for hospitality. The work of L.F.A. was supported in part by U.S.~Department of Energy
grant \#DE-FG02-90ER40542. This work was supported by the Volkswagen-Foundation.

\appendix
\section{Feynman rules for Higgsed \texorpdfstring{${\cal N}=4$}{N=4} SYM}\label{app-feynmanrules}
\label{sect-propagators}
Here we give a short list of the Feynman rules necessary for the computation in appendix \ref{appendix-oneloop}. 
Keeping the ten dimensional notation for the spinors we have to impose both the Majorana and the Weyl condition. This leads to the appearance of the ten dimensional charge conjugation matrix $C$ and the projection matrix $L=\tfrac{1}{2}(1+\Gamma^{11})$ in the Feynman rules.

\subsubsection*{Gluon propagators}
\begin{align}
\raisebox{-0.3cm}{\includegraphics[width=5cm]{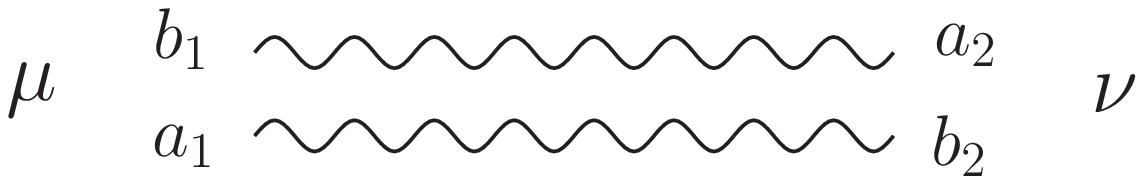}}&=\frac{-i\eta^{\mu\nu}}{q^2}\delta^{b_1}_{a_2}\delta^{b_2}_{a_1}\\
\raisebox{-0.3cm}{\includegraphics[width=5cm]{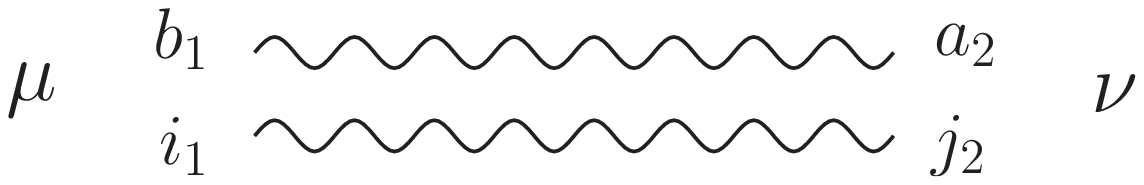}}&=\frac{-i\eta^{\mu\nu}}{q^2+m_{i_1}^2}\delta^{b_1}_{a_2}\delta^{j_2}_{i_1}\\
\raisebox{-0.3cm}{\includegraphics[width=5cm]{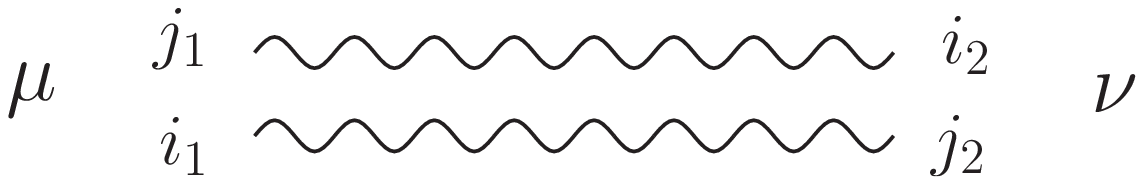}}&=\frac{-i\eta^{\mu\nu}}{q^2+(m_{i_2}-m_{i_1})^2}\delta^{j_1}_{i_2}\delta^{j_2}_{i_1}
\end{align}
\subsubsection*{Scalar propagators}
\begin{align}
\raisebox{-0.3cm}{\includegraphics[width=5cm]{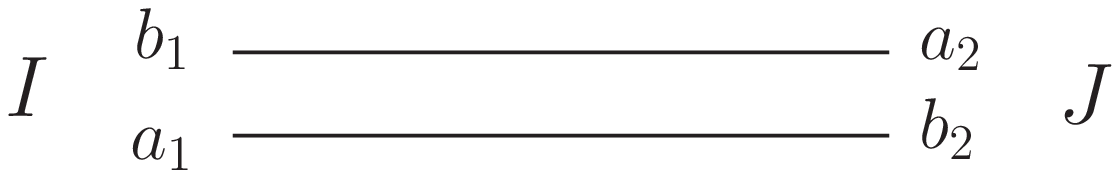}}&=\frac{-i\delta^{IJ}}{q^2}\delta^{b_1}_{a_2}\delta^{b_2}_{a_1}\\
\raisebox{-0.3cm}{\includegraphics[width=5cm]{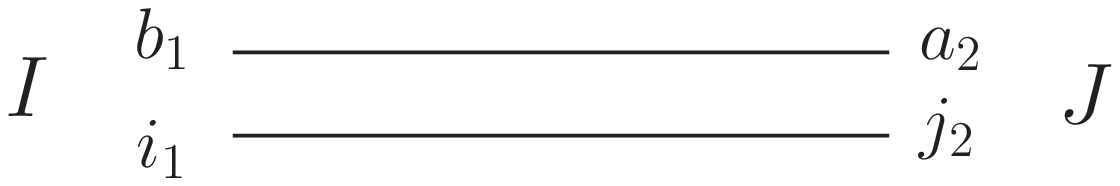}}&=\frac{-i\delta^{IJ}}{q^2+m_{i_1}^2}\delta^{b_1}_{a_2}\delta^{j_2}_{i_1}\\
\raisebox{-0.3cm}{\includegraphics[width=5cm]{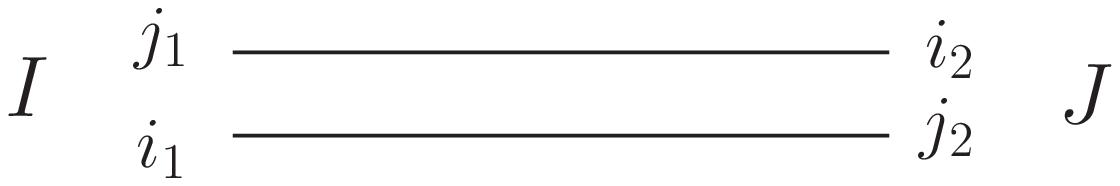}}&=\frac{-i\delta^{IJ}}{q^2+(m_{i_2}-m_{i_1})^2}\delta^{j_1}_{i_2}\delta^{j_2}_{i_1}
\end{align}
We use dotted double lines to denote the field $\Phi_{9}$.
\subsubsection*{Fermion propagators}
\begin{align}
\raisebox{-0.3cm}{\includegraphics[width=5cm]{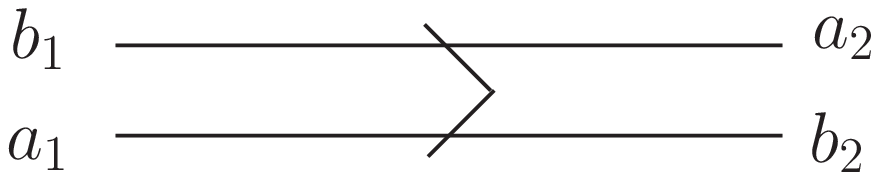}}&=\frac{iL\slashed{q}C^{-1}}{q^2}\delta^{b_1}_{a_2}\delta^{b_2}_{a_1}\\
\raisebox{-0.3cm}{\includegraphics[width=5cm]{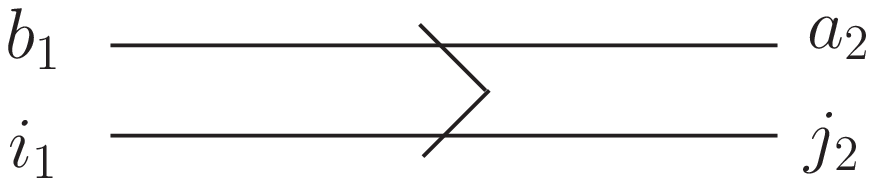}}&=\frac{iL(\slashed{q}+m_{i_1}\Gamma^9)C^{-1}}{q^2+m_{i_1}^2}\delta^{b_1}_{a_2}\delta^{j_2}_{i_1}\\
\raisebox{-0.3cm}{\includegraphics[width=5cm]{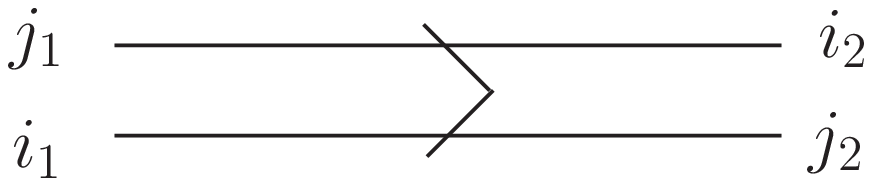}}&=\frac{iL(\slashed{q}+(m_{i_2}-m_{i_1})\Gamma^9)C^{-1}}{q^2+(m_{i_2}-m_{i_1})^2}\delta^{j_1}_{i_2}\delta^{j_2}_{i_1}
\end{align}
\subsubsection*{Vertices}
{\allowdisplaybreaks
\begin{align}
\raisebox{-2cm}{\includegraphics[width=5cm]{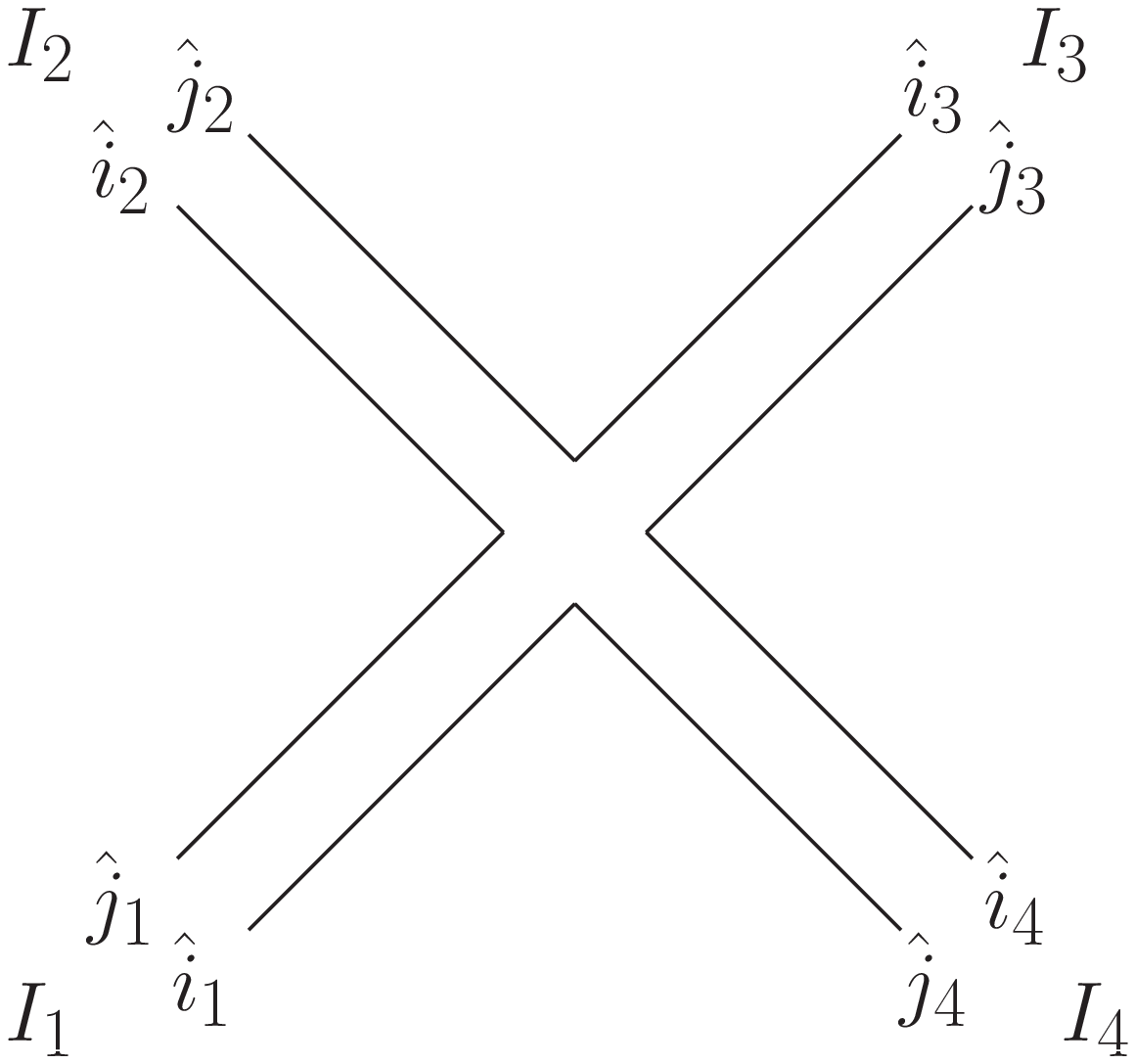}}&=\begin{aligned}[t]
                                            &ig^2\bigl({}2\delta^{I_1 I_3}\delta^{I_2 I_4}\\
&\phantom{ig^2\bigl(}-\delta^{I_1 I_2}\delta^{I_3 I_4}\\
&\phantom{ig^2\bigl(}-\delta^{I_1 I_4}\delta^{I_2 I_3}\bigr)\delta^{\hat{j}_1}_{\hat{i}_2}\delta^{\hat{j}_2}_{\hat{i}_3}\delta^{\hat{j}_3}_{\hat{i}_4}\delta^{\hat{j}_4}_{\hat{i}_1}
                                           \end{aligned}
\end{align}
\begin{align}
\raisebox{-1.5cm}{\includegraphics[width=4.5cm]{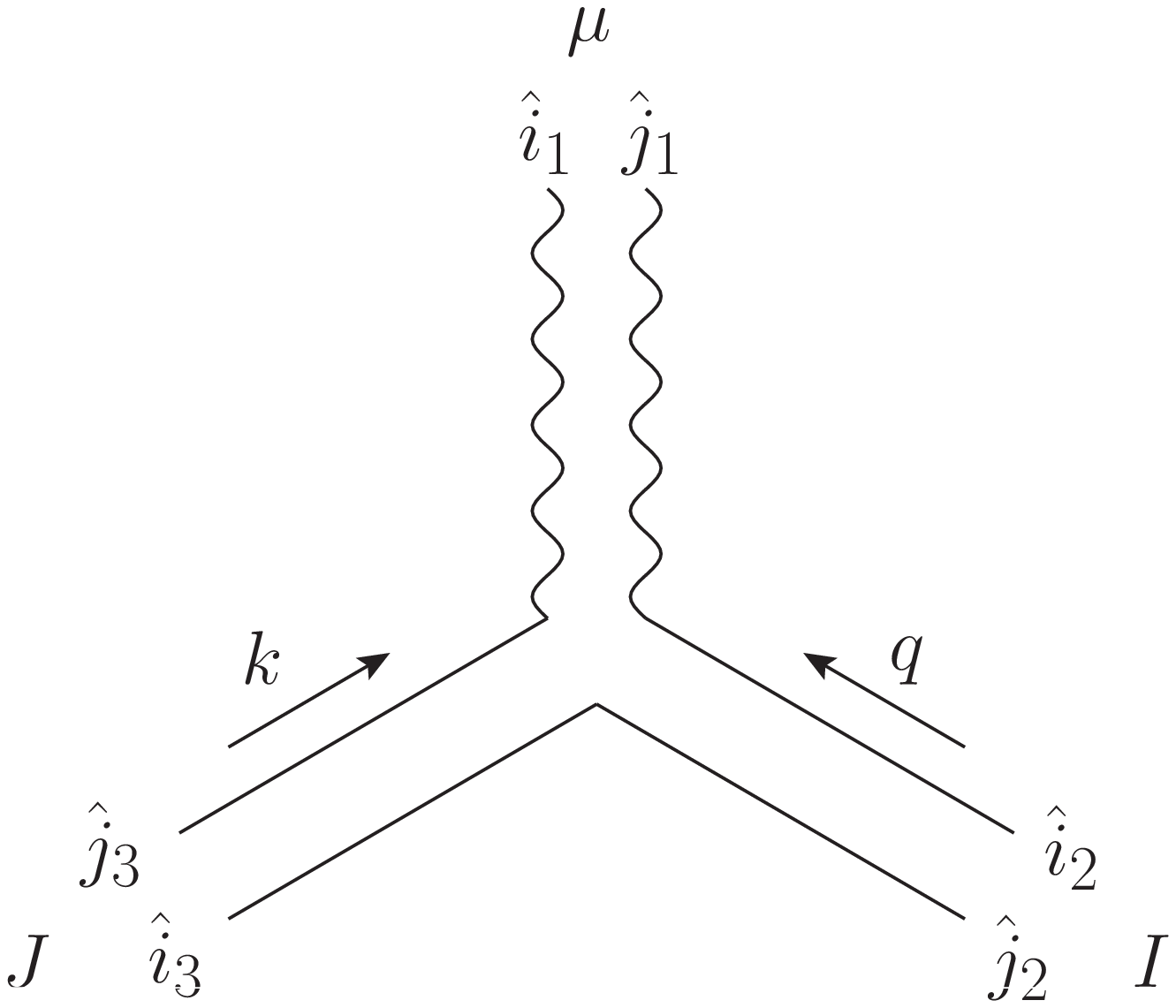}}\!\!\!\!\!\!\!\!\!\!\!\!\!&=ig(q\!-\!k)^\mu\,\delta^{IJ}\,\delta^{\hat{j}_1}_{\hat{i}_2}\delta^{\hat{j}_2}_{\hat{i}_3}\delta^{\hat{j}_3}_{\hat{i}_1}&
\raisebox{-1.5cm}{\includegraphics[width=4.5cm]{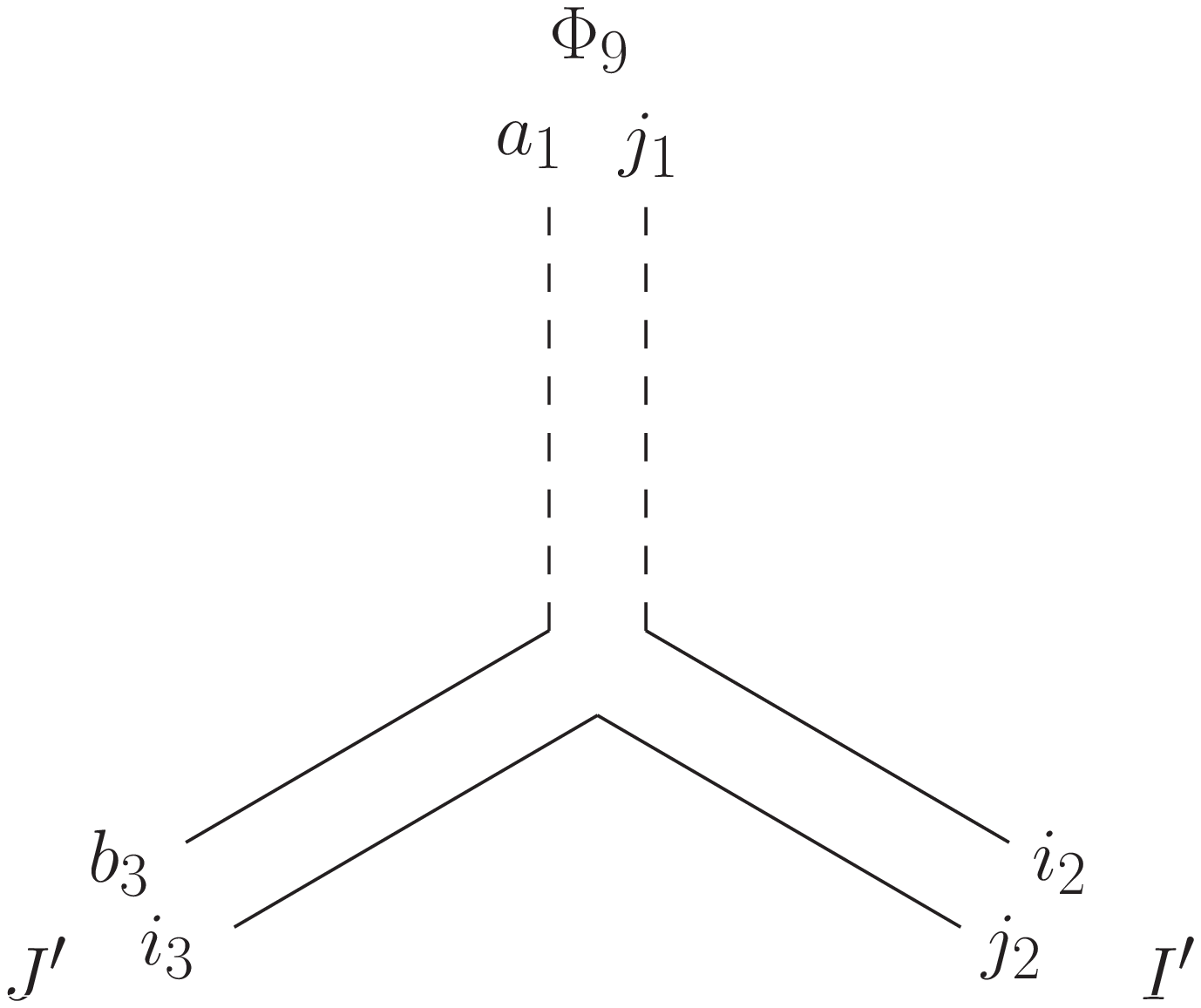}}\!\!\!\!\!\!\!\!\!\!\!\!\!\!\!\!&=ig\delta^{I'J'}\,\left(2m_{i_3}\!\!-\!m_{i_2}\right)\delta^{j_1}_{i_2}\delta^{j_2}_{i_3}\delta^{b_3}_{a_1}\\
\raisebox{-1.5cm}{\includegraphics[width=4cm]{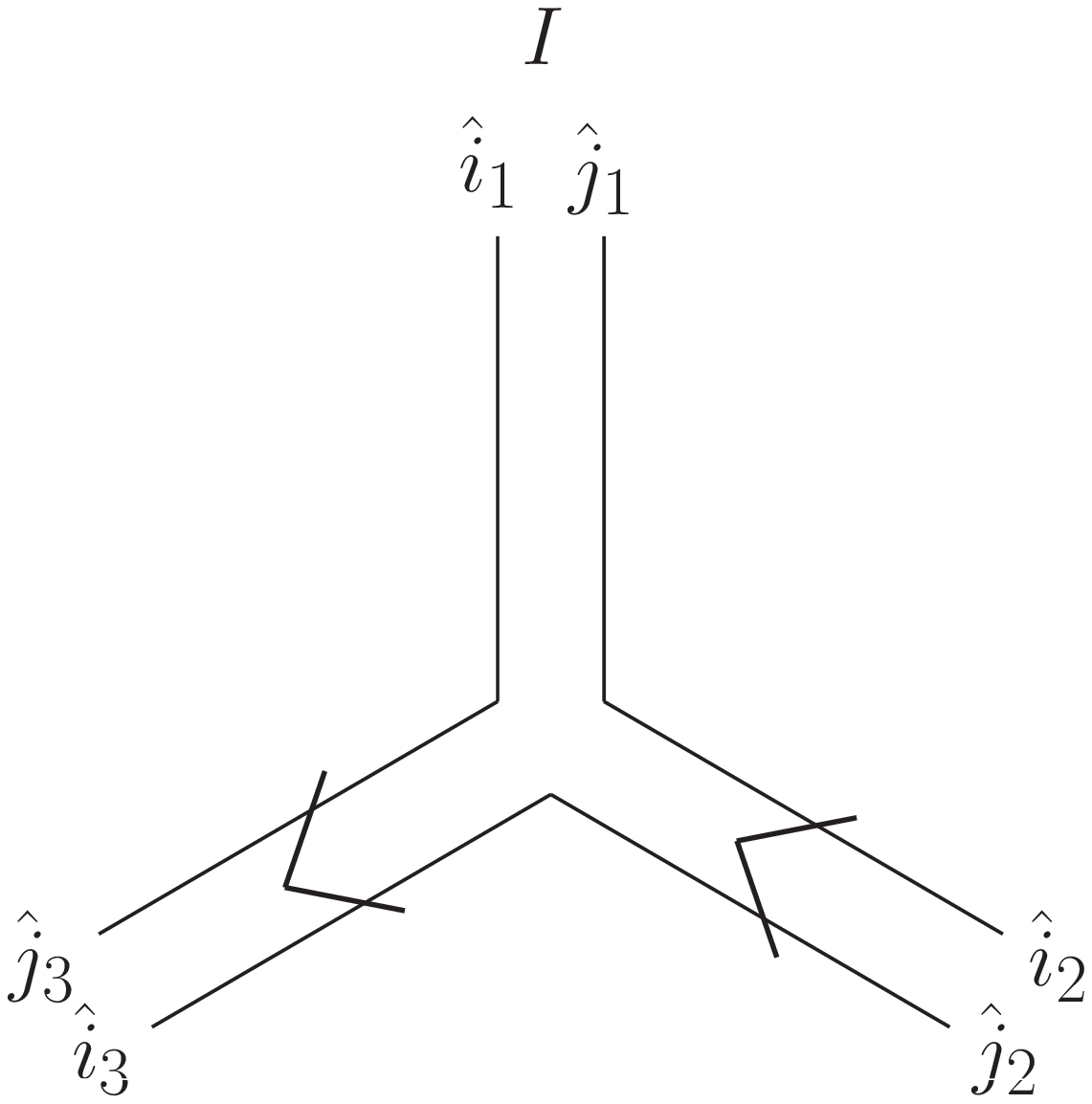}}\!\!\!\!\!\!\!\!\!\!\!\!\!&=ig\,C\Gamma^I L\,\delta^{\hat{j}_1}_{\hat{i}_2}\delta^{\hat{j}_2}_{\hat{i}_3}\delta^{\hat{j}_3}_{\hat{i}_1}&
\raisebox{-1.5cm}{\includegraphics[width=4.5cm]{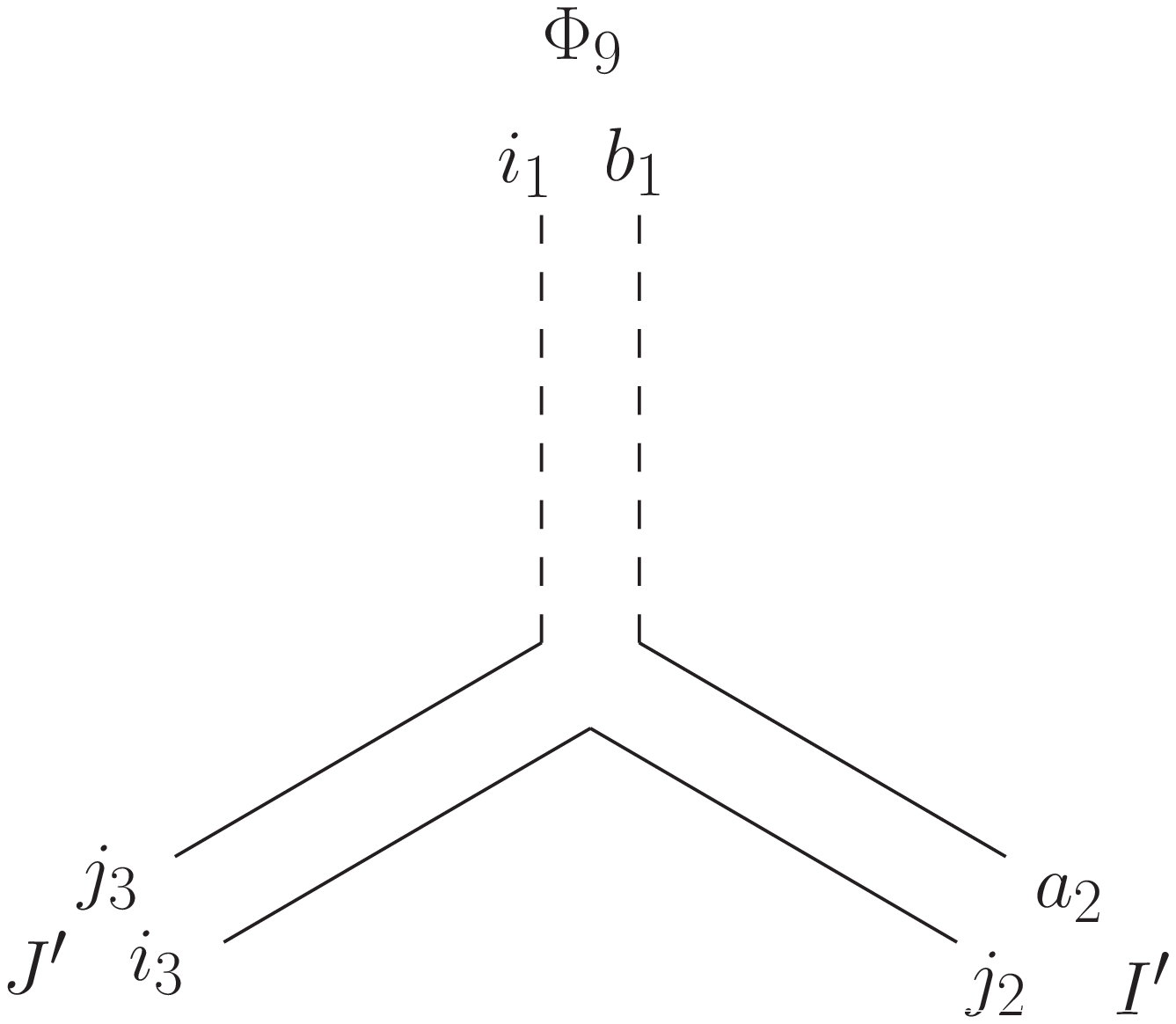}}\!\!\!\!\!\!\!\!\!\!\!\!\!\!\!\!&=ig\delta^{I'J'}\,\left(2m_{i_3}\!\!-\!m_{i_1}\right)\delta^{b_1}_{a_2}\delta^{j_2}_{i_3}\delta^{j_3}_{i_1}
\end{align}}
Hatted indices $\hat{i}$ mean that $\hat{i} \in \{1,2,...,N+M \}$.

\section{One-loop gauge theory computation}
\label{appendix-oneloop}

We want to calculate the one-loop contribution to the colour ordered amplitude
\begin{equation}
A_{4} = \langle \Phi_{I'}(p_{1}) \, \Phi_{J'}(p_{2}) \, \Phi_{I'}(p_{3}) \,\Phi_{J'}(p_{4})  \rangle\,,
\end{equation}
where we recall that $I',J' \in \{4,\ldots, 8\}$, and here we take $I'\neq J'$ with no sum on either index. We know that the amplitude is UV finite. Hence all bubble and tadpole integrals have to cancel and we can drop all bubble and tadpole diagrams from the beginning. What we are left with are the 8 triangle diagrams and the one box diagram listed in figure \ref{diagrams}. However we only need to calculate the following two triangle diagrams and obtain the others by cyclicly permuting the indices. Using the Feynman rules of appendix \ref{app-feynmanrules} we obtain:
\begin{figure}%
\center
\includegraphics[height=3cm]{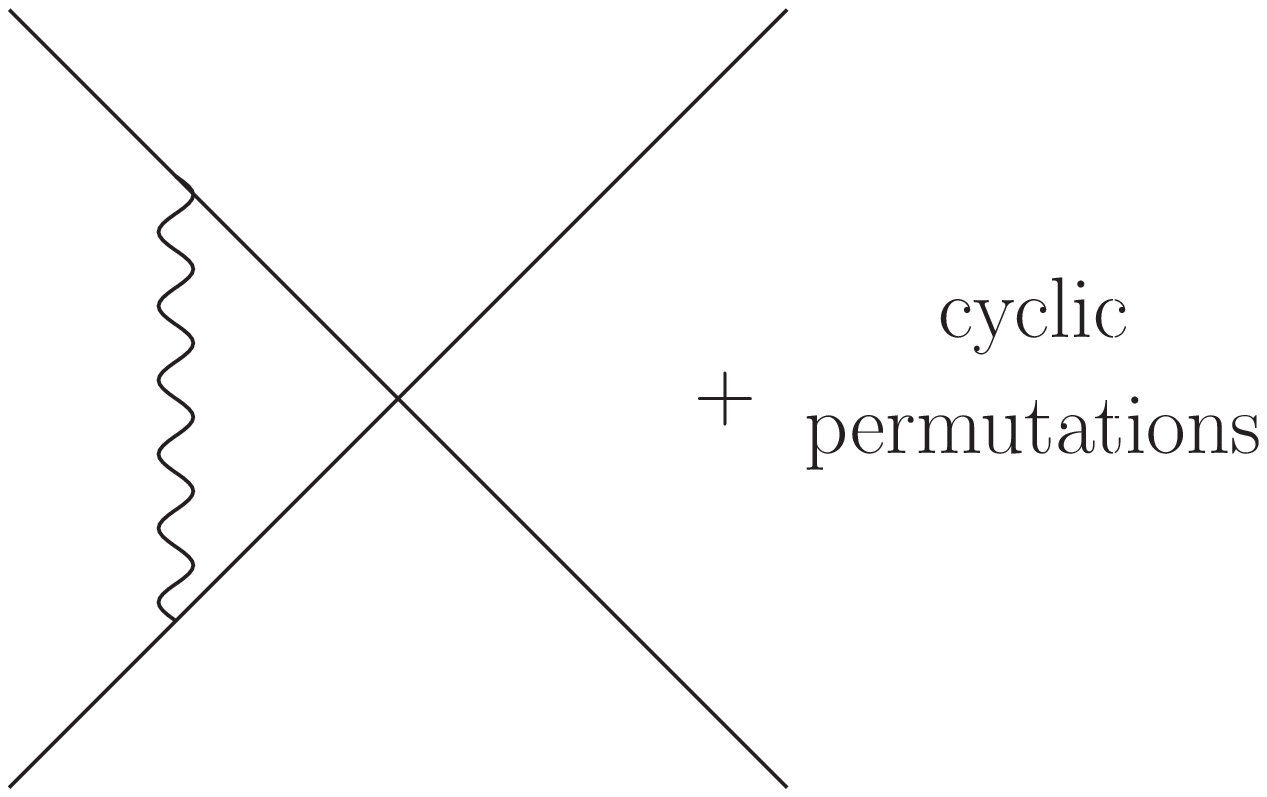}\hspace{1.5cm}\includegraphics[height=3cm]{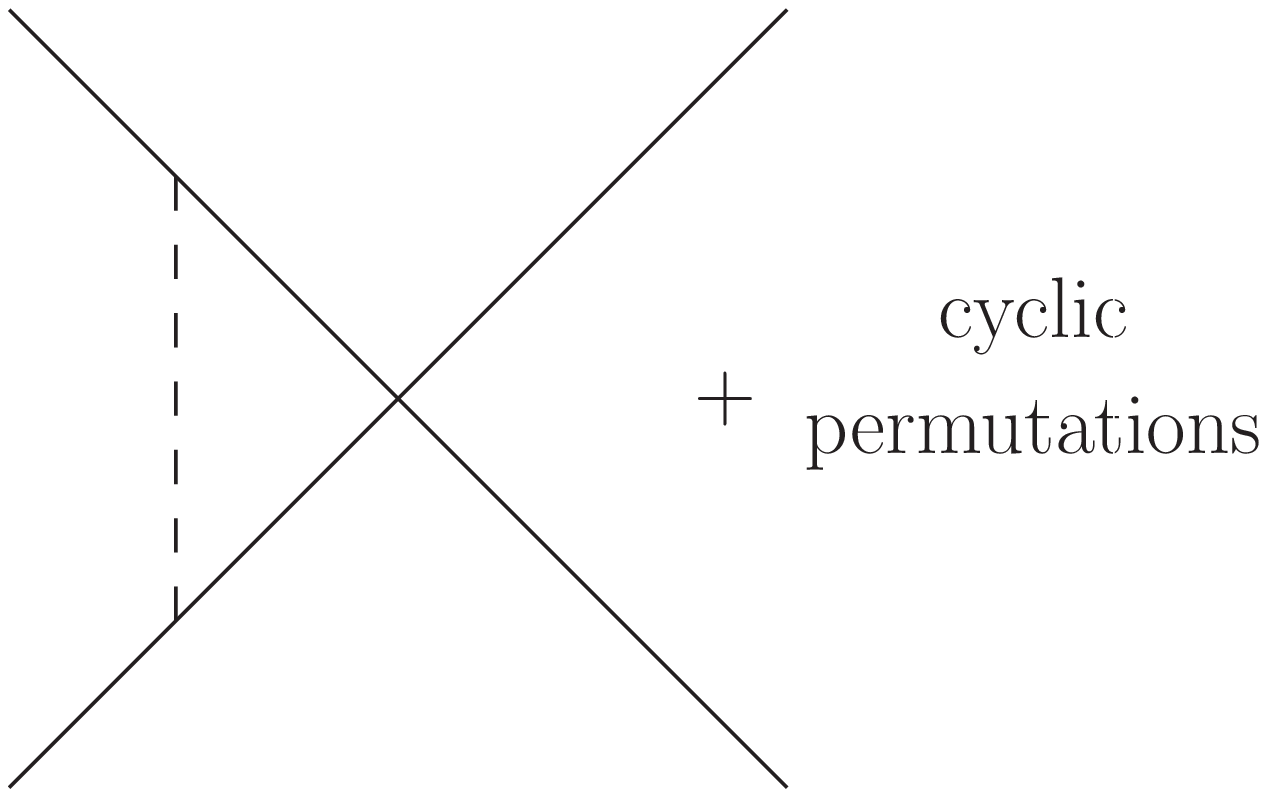}\hspace{1.5cm}\includegraphics[height=3cm]{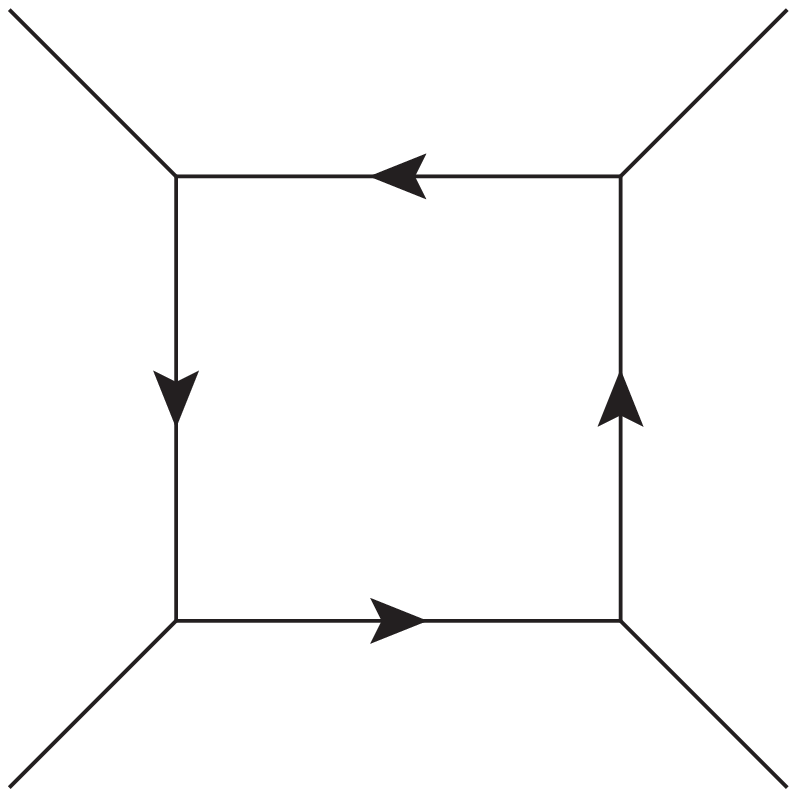}%
\caption{Relevant one-loop diagrams.}%
\label{diagrams}%
\end{figure}
{\allowdisplaybreaks
\begin{align}\label{eq:diagram1}
\raisebox{-1.5cm}{\includegraphics[width=4cm]{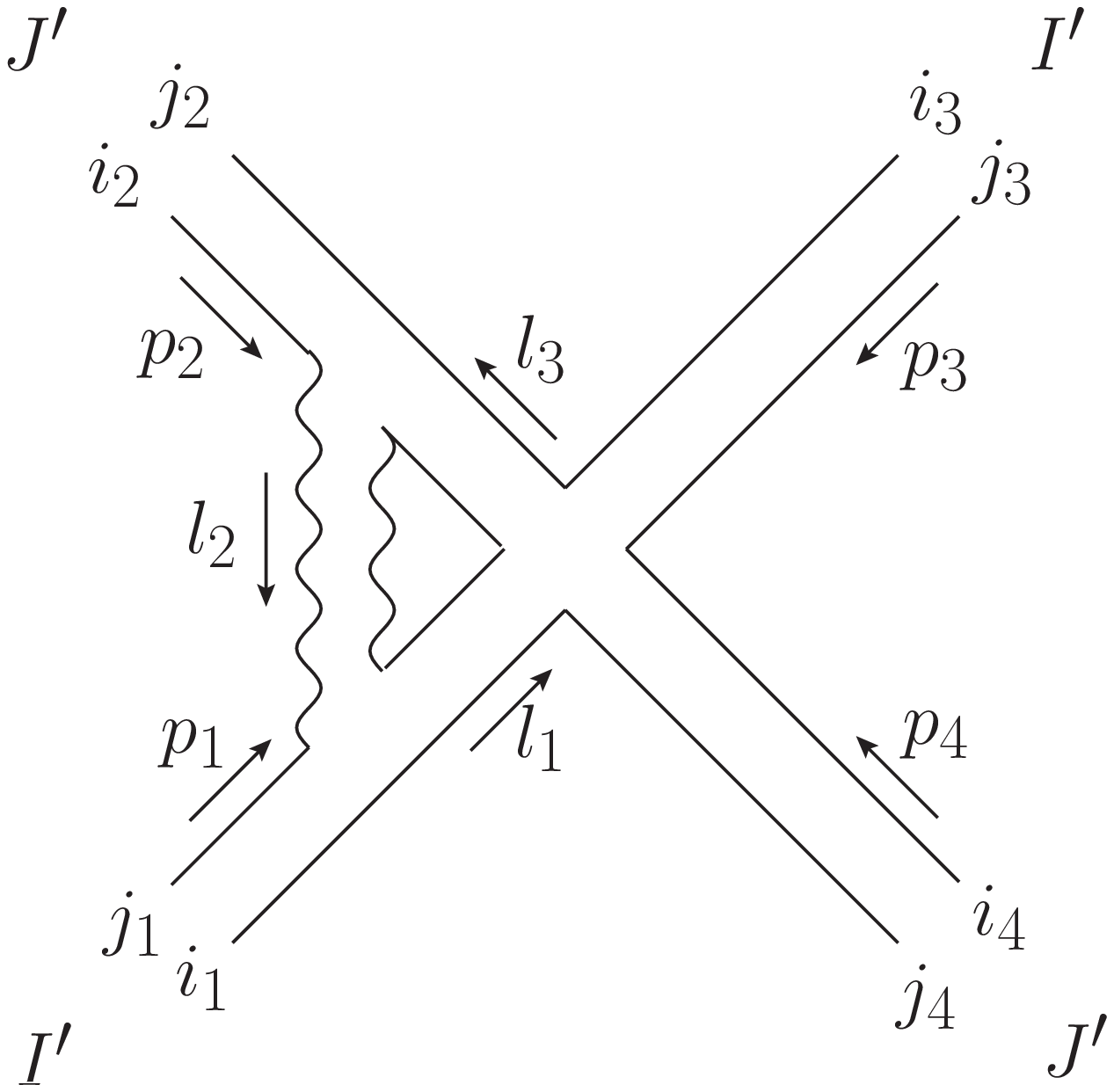}}&=2Ng^4\delta^{j_1}_{i_2}\delta^{j_2}_{i_3}\delta^{j_3}_{i_4}\delta^{j_4}_{i_1}\int\frac{d^4 l}{(2\pi)^4}\frac{(l_1+p_1)\cdot(l_3-p_2)}{(l_1^2+m_{i_1}^2)(l_2^2+m_{i_2}^2)(l_3^2+m_{i_3}^2)}\,,\\
\raisebox{-1.5cm}{\includegraphics[width=4cm]{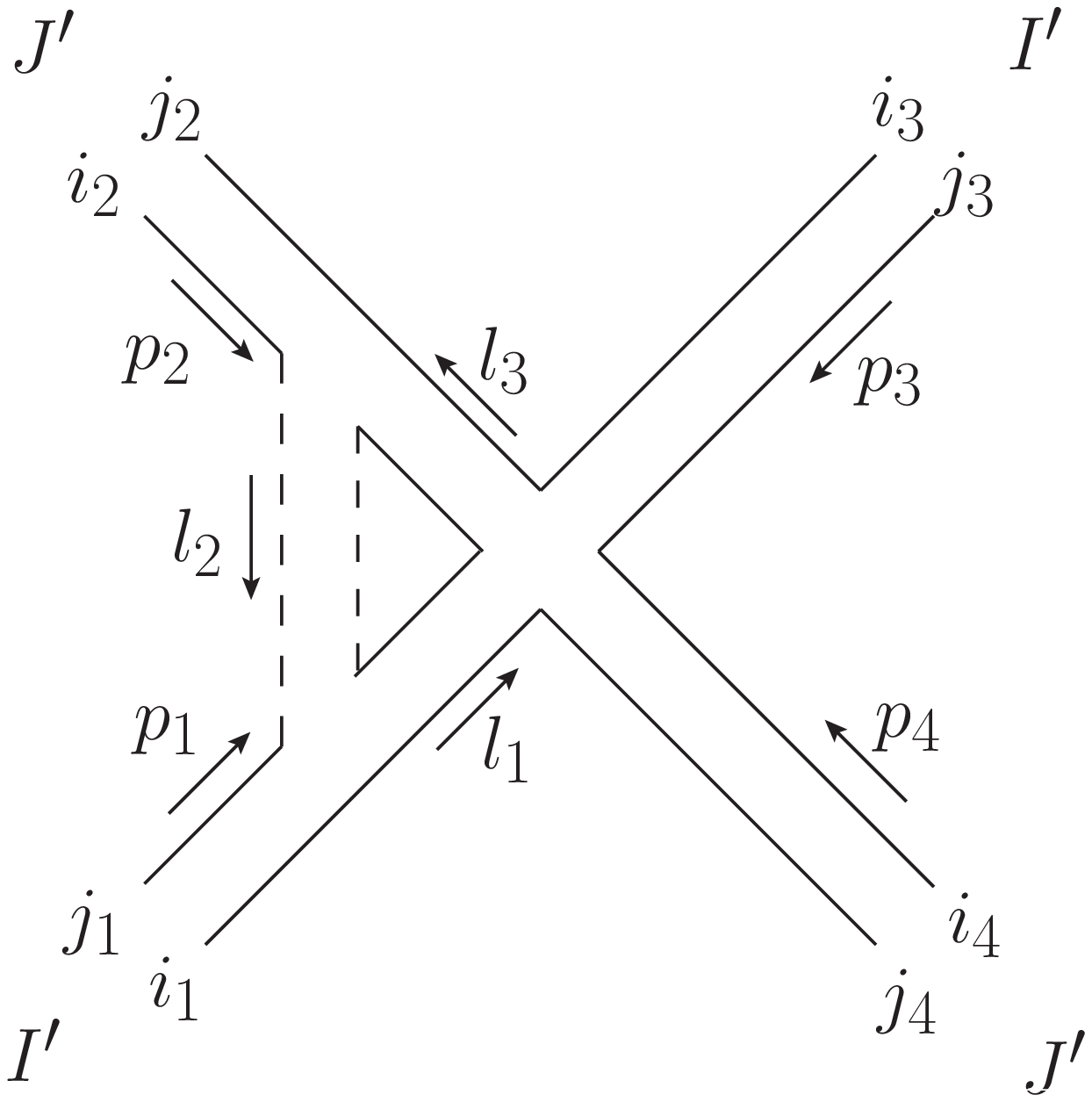}}&=2Ng^4\delta^{j_1}_{i_2}\delta^{j_2}_{i_3}\delta^{j_3}_{i_4}\delta^{j_4}_{i_1}\int\frac{d^4 l}{(2\pi)^4}\frac{(2m_{i_1}-m_{i_2})(2m_{i_3}-m_{i_2})}{(l_1^2+m_{i_1}^2)(l_2^2+m_{i_2}^2)(l_3^2+m_{i_3}^2)}\,.\label{eq:diagram2}
\end{align}}
Both diagrams can be combined using the five dimensional momenta
\begin{align}
\hat{p}_k&=(p_k,m_{i_k}-m_{i_{k+1}})&\hat{l}_k&=(l_k,m_{i_k})\,,
\end{align}
leading to
\begin{equation}
\eqref{eq:diagram1}+\eqref{eq:diagram2}=2Ng^4\delta^{j_1}_{i_2}\delta^{j_2}_{i_3}\delta^{j_3}_{i_4}\delta^{j_4}_{i_1}\int\frac{d^4 l}{(2\pi)^4}\frac{(\hat{l}_2+2\hat{p}_1) \cdot (\hat{l}_2-2\hat{p}_2)}{\hat{l}_1^2\,\hat{l}_2^2\,\hat{l}_3^2}
\end{equation}
Because of the two identities
\begin{align}
2\hat{l}_2 \cdot \hat{p}_2&=\hat{l}_2^2-\hat{l}_3^2&\text{and} && 2\hat{l}_2 \cdot \hat{p}_2&=\hat{l}_1^2-\hat{l}_2^2
\end{align}
the triangle coefficient is simply $-8Ng^4\hat{p}_1\cdot \hat{p}_2$. Summing up all triangle diagrams we obtain
\begin{equation}
-8Ng^4\delta^{j_1}_{i_2}\delta^{j_2}_{i_3}\delta^{j_3}_{i_4}\delta^{j_4}_{i_1}\int\frac{d^4 l}{(2\pi)^4}\left(\frac{\hat{p}_1 \cdot \hat{p}_2}{\hat{l}_1^2\,\hat{l}_2^2\,\hat{l}_3^2}+\frac{\hat{p}_2 \cdot \hat{p}_3}{\hat{l}_2^2\,\hat{l}_3^2\,\hat{l}_4^2}+\frac{\hat{p}_1 \cdot \hat{p}_2}{\hat{l}_3^2\,\hat{l}_4^2\,\hat{l}_1^2}+\frac{\hat{p}_2 \cdot \hat{p}_3}{\hat{l}_4^2\,\hat{l}_1^2\,\hat{l}_2^2}\right)\quad+\text{bubbles}\,.
\label{eq:triangles}
\end{equation}
The box diagram is given by
\begin{align}
&\raisebox{-1.7cm}{\includegraphics[width=4.5cm]{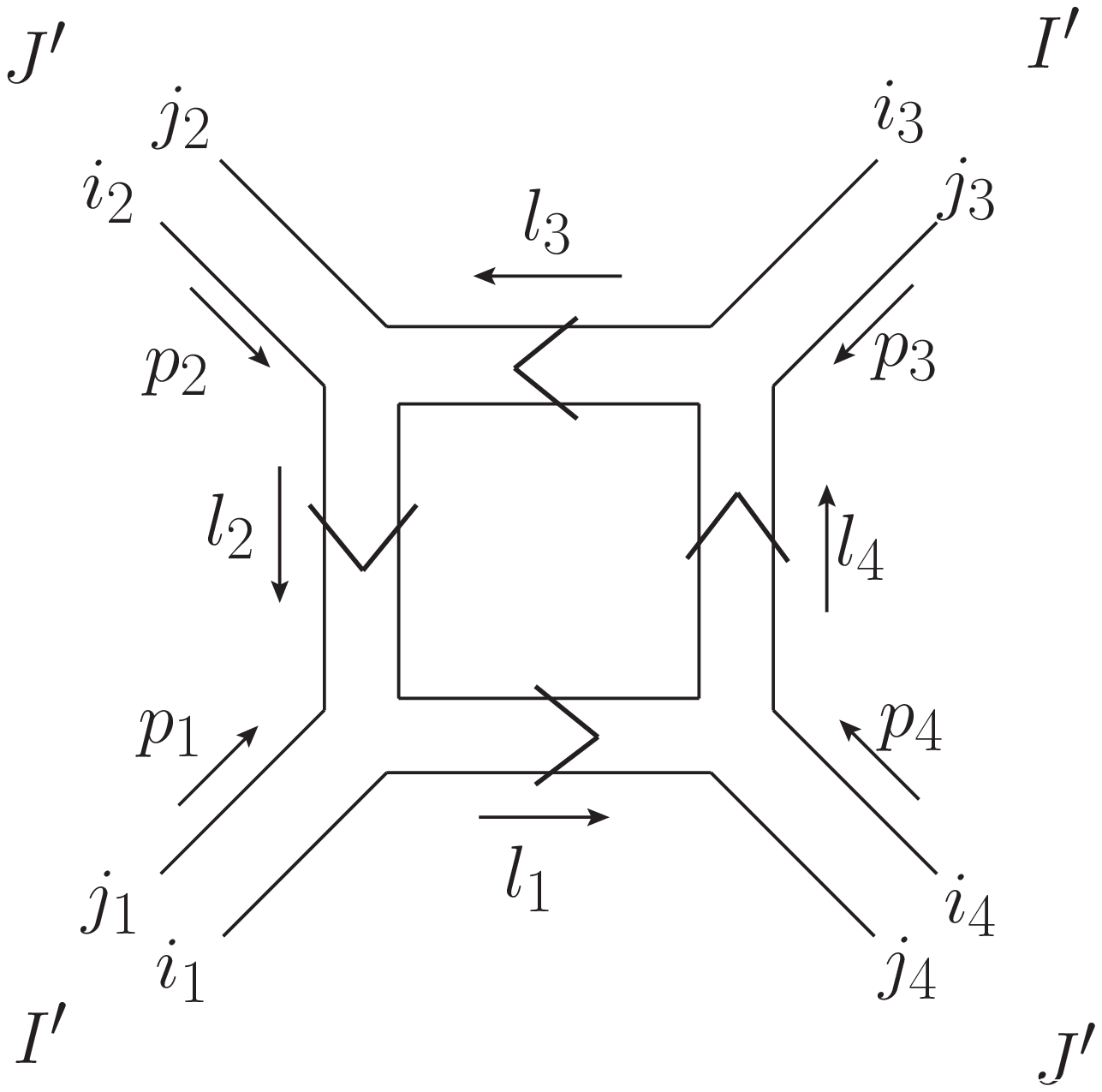}}\begin{aligned}&=-Ng^4\delta^{j_1}_{i_2}\delta^{j_2}_{i_3}\delta^{j_3}_{i_4}\delta^{j_4}_{i_1}\int\frac{d^4 l}{(2\pi)^4}\frac{\Tr\left[\Gamma^{I'}L\hat{\slashed{l}}_1\Gamma^{J'}L\hat{\slashed{l}}_4\Gamma^{I'}L\hat{\slashed{l}}_3\Gamma^{J'}L\hat{\slashed{l}}_2\right]}{\hat{l}_1^2\,\hat{l}_2^2\,\hat{l}_3^2\,\hat{l}_4^2}\\
&=Ng^4\delta^{j_1}_{i_2}\delta^{j_2}_{i_3}\delta^{j_3}_{i_4}\delta^{j_4}_{i_1}\int\frac{d^4 l}{(2\pi)^4}\frac{\Tr\left[L\hat{\slashed{l}}_1\,\hat{\slashed{l}}_4\,\hat{\slashed{l}}_3\,\hat{\slashed{l}}_2\right]}{\hat{l}_1^2\,\hat{l}_2^2\,\hat{l}_3^2\,\hat{l}_4^2}\end{aligned}\notag\\
&=16Ng^4\delta^{j_1}_{i_2}\delta^{j_2}_{i_3}\delta^{j_3}_{i_4}\delta^{j_4}_{i_1}\int\frac{d^4 l}{(2\pi)^4}\frac{(\hat{l}_1 \cdot \hat{l}_4)(\hat{l}_3 \cdot \hat{l}_2)-(\hat{l}_1 \cdot \hat{l}_3)(\hat{l}_2 \cdot \hat{l}_4)+(\hat{l}_1 \cdot \hat{l}_2)(\hat{l}_4 \cdot \hat{l}_3)}{\hat{l}_1^2\,\hat{l}_2^2\,\hat{l}_3^2\,\hat{l}_4^2}\,.
\label{eq:diagram3}
\end{align}
Only the term $(\hat{l}_1 \cdot \hat{l}_3)(\hat{l}_2 \cdot \hat{l}_4)$ contributes to the box and triangle integrals. Making use of the simple identity
\begin{equation}
(\hat{l}_1 \cdot \hat{l}_3)(\hat{l}_2 \cdot \hat{l}_4)=\left(\tfrac{1}{2}\hat{l}_1^2+\tfrac{1}{2}\hat{l}_3^2-\hat{p}_1 \cdot \hat{p}_2\right)\left(\tfrac{1}{2}\hat{l}_2^2+\tfrac{1}{2}\hat{l}_4^2-\hat{p}_2 \cdot \hat{p}_3\right)\,,
\end{equation}
we can read off the box and triangle coefficients and see that the triangle integrals indeed cancel:
\begin{equation}
 \eqref{eq:diagram3}+\eqref{eq:triangles}=-16Ng^4\delta^{j_1}_{i_2}\delta^{j_2}_{i_3}\delta^{j_3}_{i_4}\delta^{j_4}_{i_1}\int\frac{d^4 l}{(2\pi)^4}\frac{(\hat{p}_1 \cdot \hat{p}_2)(\hat{p}_2 \cdot \hat{p}_3)}{\hat{l}_1^2\,\hat{l}_2^2\,\hat{l}_3^2\,\hat{l}_4^2}\quad+\text{bubbles}\;.
\label{app-oneloopfeynman}
\end{equation}

\section{Results for integrals in the Higgsed theory}
\label{appendix-integrals}

\subsection{One-loop box integral by Mellin-Barnes method}
We illustrate the use of the Mellin-Barnes (MB) representation for computing
loop integrals (for further reading see \cite{Smirnov:2004ym}). Consider the one-loop box integral of equation (\ref{int0}) with massless external lines
and a uniform mass $m_{i}=m$ circulating in the loop. Using the Mathematica
package {\it AMBRE} \cite{Gluza:2007rt}, a two-fold MB representation is automatically generated.
It reads
\begin{align}
I^{(1)}(s,t,m)  &=  \int d\tilde{z}_{1} d\tilde{z}_{2}
(m^2)^{z_{1}}
s^{(1+z_{2})}
 t^{(-1-z_{1}-z_{2})}  \times \nonumber \\
&\=  \Gamma(-z_{1})
 \Gamma^2(-1 -z_1 -z_2 )
 \Gamma(-z_2 ) \Gamma^2(1+z_2 ) \Gamma(2+z_1 + z_2 )
\frac{1}{\Gamma(-2 z_1 )} \,,
\end{align}
where $d\tilde{z} = dz/(2 \pi i)$ and the real part of $z_{1},z_{2}$ can be taken to be $-1$ and $-1/2$, respectively.
In order to take the small $m$ limit, we want to deform the integration contour for $z_{1}$ such that its real part is positive.
In doing so, we pick up a pole at $z_{1}=-1-z_{2}$ originating from $ \Gamma^2(-1 -z_1 -z_2 )$.
Note that the pole at $z_{1}=0$ is spurious.
The deformed integral with $Re(z_{1})>0$ vanishes as $m^2 \to 0$, therefore, taking the residue at $z_{1}=-1-z_{2}$ we obtain
\begin{align}
I^{(1)}(s,t,m) &= -\int  d\tilde{z}_{2} (m^2)^{(-1-z_{2})} s^{(1+z_{2})}  \frac{\Gamma(-z_2 ) \Gamma^3(1+z_2 )}{\Gamma(2 + 2 z_2 )} \nonumber \\
&\=\phantom{-\int  d\tilde{z}_{2} } \times \left[ -h(z_{2}) + 2 h(1 + 2 z_{2}) + \ln(m^2/t) \right]+ O(m^2)\,,
\end{align}
where $Re(z_{2}) =-1/2$ and $h(z) = \Psi(1+z) - \gamma$.
We can reiterate the above procedure and deform the integration contour for $z_{2}$. We want to take it from  $Re(z_{2}) =-1/2$ to
 $Re(z_{2}) <-1$. In doing so, we pick up a pole at $Re(z_{2}) = -1$.
Taking the residue, we obtain the final result
\begin{equation}
I^{(1)}(s,t,m) = 2 \ln(m^2/s) \ln(m^2/t) - \pi^2 + O(m^2) \,.
\end{equation}
Note that in this simple example we were able to find the answer without doing any integrations,
just by using Cauchy's theorem.

\subsection{Two-loop box integral by Mellin-Barnes method}
The Mathematica package {\it AMBRE} \cite{Gluza:2007rt} automatically produces the following five-fold MB representation
for the two-loop box integral of equation (\ref{doubleboxequalmass}),
\begin{align}
{I}^{(2)}(s,t,m) &= \int d\tilde{z}_{1} \, d\tilde{z}_{2} \, d\tilde{z}_{3} \, d\tilde{z}_{7} \, d\tilde{z}_{8} \, (m^2)^{(z_{1} + z_{7})} \,
s^{(1 - z_{1} + z_{8})} \, t^{(-1 - z_{7} - z_{8})} \nonumber \\
 & \=\times   \Gamma(-z_{1}) \Gamma(1 + z_{1}) \Gamma(-z_{2}) \Gamma(-z_{1} + z_{2}) \Gamma(-z_{2} -
     z_{3}) \Gamma(-z_{3}) \Gamma(-z_{1} + z_{3}) \nonumber \\
    &\= \times \Gamma(-z_{7}) \Gamma(-1 - z_{7} -
     z_{8}) \Gamma(-1 + z_{2} + z_{3} - z_{7} - z_{8}) \Gamma(-z_{8})  \nonumber \\
    &\= \times \frac{\Gamma(
    1 - z_{2} + z_{8}) \Gamma(1 - z_{3} + z_{8}) \Gamma(
    2 + z_{7} + z_{8}) }{\Gamma(-2 z_{1}) \Gamma(1 - z_{2}) \Gamma(
    1 - z_{3}) \Gamma(-2 z_{7} ) }\,.
\end{align}
Here all integrations go from $-i \infty$ to $+i \infty$ and the real part of the integration variables $z_{1}, z_{2}, z_{3}, z_{7}, z_{8}$ is
taken to be $-21/32, -1/8,  -1/4,  -7/8, -9/16$ , respectively. Although this formula may appear somewhat
complicated at first glance, it is very easy to extract the small $m^2$ expansion from it, just as in the one-loop example of the previous subsection.
One obtains a few constant two- and one-fold MB integrals, and only one simple one-fold MB integral that depends on $x=s/t$, namely
\begin{align}
f(x) &:= \frac{1}{2 \pi i} \int_{1/2-i \infty}^{1/2+i \infty} x^z \Gamma^3(-z) \Gamma^2(z) \Gamma(1+z) dz \nonumber \\
&\phantom{:}=   \frac{1}{2} \left[ \pi^2 {\rm Li}_{2}(-x)  + \ln^2(x) {\rm Li}_{2}(-x) -4 \ln(x) {\rm Li}_{3}(-x) + 6 {\rm Li}_{4}(-x) \right]\,.
\end{align}
All other contributions are obtained via Cauchy's theorem without doing any integrations.
The result is
\begin{align} \label{finaldoublebox}
{I}^{(2)}(s,t,m) &= \frac{1}{3} \ln^{4}(u)  - \frac{4}{3} \ln^3(u) \ln(v)   + 2 \ln^2(u) \ln^2(v) +
2 \pi^2 \ln^2(u)  - \frac{8}{3} \pi^2 \ln(u) \ln(v)  \nonumber \\
&\= -4 \zeta_{3} \ln(v)
 + 8 f(v/u) + \frac{2}{3}\pi^4
 + O(m^2) \,,
\end{align}
where $u=m^2/s$ and $v=m^2/t$.
A short calculation gives
\begin{multline}\label{appendix-I2stsymm}
\frac{1}{4} {I}^{(2)}(s,t,m) + \frac{1}{4} {I}^{(2)}(t,s,m) = \frac{1}{4} \left[ \ln^4(u) + \ln^4(v) \right] + \left[ -\frac{1}{2} \ln^2(v/u)- \zeta_{2}\right] \left[ \ln^2(u) + \ln^2(v) \right]  \\
 - \zeta_{3}  \left[ \ln(u) + \ln(v) \right] + \left[   \frac{1}{4} \ln^4(v/u) +  \zeta_{2} \ln^2(v/u)+\frac{1}{20} \pi^4  \right] +O(m^2)\,.
\end{multline}
Finally, we also compute the square of the one-loop result that is needed in order to check the exponentiation at two loops, see equation (\ref{wtwoloops}),
\begin{align}\label{appendix-I1squared}
-\frac{1}{8} (I^{(1)}(s,t,m))^2 &= - \frac{1}{4}  \left[ \ln^4(u) + \ln^4(v) \right] +\left[ \frac{1}{2} \ln^2(v/u)  +\frac{1}{4} \pi^2 \right] \left[ \ln^2(u) + \ln^2(v) \right]  \nonumber \\
&\= - \frac{1}{4} \ln^4(v/u)  - \frac{1}{4}  \pi^2 \ln^2(v/u)  - \frac{1}{8} \pi^4 +O(m^2) \,.
\end{align}

\subsection{Generic dual conformal one-loop box integrals}
Here we give the one-loop scalar dual conformal box integrals that appear in the Higgsed theory.
A generic one-loop dual conformal integral  is given by
\begin{equation}
 {J}^{(1)}(\hat{x}_{r},\hat{x}_{s},\hat{x}_{t},\hat{x}_{u}) =  c_{0} \, \hat{x}^2_{rt} \hat{x}^2_{su} \int d^{5}\hat{x}_{0} \frac{\delta(\hat{x}_{0}^{M=4})}{ \hat{x}_{r0}^2 \hat{x}_{s0}^2 \hat{x}_{t0}^2 \hat{x}_{u0}^2}\,,
\end{equation}
in dual notation. This integral generalises the $4$-point dual conformal integral given in
section \ref{S-one-loop} to an arbitrary number of points.
Equivalently, in momentum notation (setting with $m_{i}=m$ for convenience) we have
\begin{align}\label{generaloneloopint}
 {J}^{(1)}(K_{1},K_{2},K_{3},K_{4},m) &=  c_{0} \, P^{-1}( K_{1}+K_{2},m^2) P^{-1}( K_{2}+K_{3},m^2) \\
 &\= \times \,  \int d^{4}k \,  P(k,m^2)  P(k+K_{1},m^2)  P(k+K_{1}+K_{2},m^2)  P(k-K_{4},m^2)  \,,\notag
\end{align}
where $P(k,m^2) = (k^2 + m^2)^{-1}$ and $K_{1} =p_{r} + \ldots + p_{s-1} , K_{2}=p_{s} + \ldots + p_{t-1},  K_{3}= p_{t} + \ldots + p_{u-1}, K_{4} =p_{u} + \ldots + p_{r-1}$,
and $p_{i}^2=0$.
The $K_{i}$ can be light-like if they are built from one momentum only. If $q$ is the number of non-light-like $K_{i}$, we call the integral in (\ref{generaloneloopint})
$q$-mass with uniform internal mass $m$, in analogy with the nomenclature for the corresponding integrals in dimensional regularisation.
The explicit expressions for ${J}^{(1)}$ can be obtained from \cite{'tHooft:1978xw}.
See figure \ref{figure-fivepoints} for the specific example of the 1-mass integral with uniform internal mass $m$.

\section{Berkovits-Maldacena solution in conformal gauge}
\label{stringapp}

The world-sheet relevant for the regularised scattering amplitudes considered in this paper ends in a light-like polygon at some finite radial distance $r_c$ from the boundary. Unfortunately, such a solution is very hard to construct. On the other hand, one could consider the simplified problem of a single cusp ending at $z=r_c$. The scattering solution should then be given by this one when approaching any of the cusps.

The solution for a single cusp was given by Berkovits and Maldacena (BM) in the appendix of \cite{Berkovits:2008ic}, as a solution of the equations of motion of the Nambu-Goto action.  The solution reads
\begin{align}
T&=e^\tau \cosh \sigma,&X&=e^\tau \sinh \sigma,&Z(\tau,\sigma)&=e^\tau \omega(\tau)
\end{align}
with
\begin{equation}
\frac{e^\tau}{r_c}=\left(\frac{w+\sqrt{2}}{w-\sqrt{2}} \right)^{\frac{1}{\sqrt{2}}}\frac{1}{1+w}
\end{equation}
The cusp is located at $\tau \rightarrow \infty$ and $\omega=r_c e^{-\tau}+1+...$. It is not possible to give a closed form for $\omega(\tau)$, however, it is very easy to solve for it as a power expansion in $e^\tau$, close to the cusp. For the first few orders we obtain
\begin{equation}
Z(\tau)=r_c+e^\tau+\frac{2}{3 r_c^2}e^{3\tau}-\frac{2}{r_c^3}e^{4\tau}+\frac{26}{5 r_c^4}e^{5\tau}+\ldots
\end{equation}
without following an apparent pattern. Notice that the above gives $Z$ as a function of $e^\tau=\sqrt{T^2-X^2}$.

\subsection{Conformal gauge}

For many purposes, a solution to the equations of motion in conformal gauge is desirable. According to the above analysis, we propose an ansatz of the form
\begin{align}
\label{bmans}
T(u,v)&=f(v) \cosh u,&X(u,v)&=f(v) \sinh u,&Z(u,v)&=g(v)
\end{align}
with boundary conditions $f(v)=0+v+...,g(v)=r_c+...$, namely, the cusp is located at $v=0$. The topology of the World-sheet is that of the upper half plane, hence equivalent to the disk. We have checked that with the above ansatz we can write a series expansion for $f(v)$ and $g(v)$ and satisfy both the equations of motion and the Virasoro constraints order by order in the $v$ expansion (this is non trivial, since there are more equations than free parameters). This of course is due to the symmetries of the problem. The Virasoro constraints are particularly simple and imply
\begin{equation}
f(v)^2+f'(v)^2-g'(v)^2=0
\end{equation}
As already mentioned, one can solve the above equations order by order in $v$, obtaining for the first few terms
\begin{equation}
f(v)= k v+\frac{k^2}{r_c}v^2+\ldots,~~~~~g(v)=r_c+k v+\frac{k^2}{r_c}v^2+\ldots
\end{equation}
At this point the solution depends on a free parameter $k$, which is not fixed by the equations of motion or Virasoro constraints. Such coefficient, presumably can be fixed by requiring the correct boundary conditions. In order to compare this solution with the BM solution we can express $Z=g(v)$ in terms of $f(v)=\sqrt{T^2-X^2}$, we obtain
\begin{equation}
g(v)=r_c+f(v)+\frac{1}{6 k^2}f(v)^3-\frac{1}{2 r_c k^2}f(v)^4+\ldots
\end{equation}
We see that the BM solution corresponds to $k=\pm r_c/2$. Setting this value, all the terms in the expansion match the corresponding terms in the  BM solution, so the solutions are indeed the same. (presumably the same value for $k$ can be found by requiring the correct boundary conditions, for instance $g(v)=\sqrt{2}f(v)$ for large $v$.)

The main lesson is that a solution can be constructed and indeed has the topology of the upper half plane, as expected for a regularised world-sheet.

\subsection{A pleasant surprise}

One advantage of writing a solution of strings on $AdS_3$ in conformal gauge, is that one can perform a Pohlmeyer type reduction as done in \cite{Alday:2009ga}. There, it was seen that given a solution of classical strings on $AdS_3$, one could obtain a holomorphic function $p(z)$ (with $z=u+i v$) plus a field $\alpha(z,\bar{z})$ (or $\hat \alpha(w,\bar{w})$) satisfying the generalised Sinh-Gordon equation.  In \cite{Alday:2009ga} it was found that for the problem relevant to scattering amplitudes, $p(z)$ is simply given by a polynomial and $\hat \alpha$ is such that it decays at infinity (with $\alpha$ regular everywhere). The area of the world-sheet is then obtained by expressing the conformal gauge action in terms of the reduced fields and is simply
\begin{equation}
A=\int e^{\hat\alpha(w,\bar w)}dw d\bar{w}
\end{equation}
One natural question is to which field and holomorphic function does the above solution correspond to. Performing the reduction we find (order by order in $v$)
\begin{align}
p(z)&=i,&e^{\hat \alpha}&=\frac{v^2}{2}-\frac{v^4}{6}+\frac{17}{360}v^6+\ldots
=\tanh^2\left(\frac{v}{\sqrt{2}} \right)
\end{align}
Namely, both quantities have a very simple expression and they can be written in a closed form! Notice that $\hat \alpha(v)$ satisfies the Sinh-Gordon equation \footnote{Actually, this is nothing but the soliton solution of Sinh-Gordon, see {\it e.g.} eq. (3.2) of \cite{Jevicki:2007aa}. However, note that the space-time interpretation of this solution (and the topology of the world-sheet) is very different from the one of that paper.} (for the particular case in which $\alpha$ depends only on $v$), namely
\begin{equation}
\hat \alpha''(v)=2 \sinh \hat \alpha(v)
\end{equation}
Also, notice that $\hat \alpha$ has the correct boundary conditions corresponding to a scattering problem, since it vanishes at $v \rightarrow \infty$ (see  \cite{Alday:2009ga} for the details).

The solution corresponding to a general scattering can then be chosen to live in the upper half plane, such that the cusps are located at $v=0$ (and each cusp corresponds to a segment). When approaching one of the cusps (and far from the others) the solution should approach the single cusp solution given here.

\subsection{Computing the leading area}

Once we have the single cusp solution, we can extract the value of the cusp anomalous dimension at strong coupling by computing the area of the corresponding world-sheet. The single cusp solution possesses both, IR and UV divergences. UV divergences (or IR depending how the solution is interpreted) are regularised by putting boundary conditions at $r=r_c$. On the other hand, we can set a IR cut-off, for instance by considering $T<t_c$. By dimensional analysis the area should then depend on the dimensionless quantity $\frac{t_c}{r_c}$. We are interested in the value of the area for large  values of $\frac{t_c}{r_c}$.

First, notice that the cusp is located at $v=0$. However, in the regularisation we are using, the contribution to the area from the $v \approx 0$ region is small (since it is finite). The biggest contribution comes from large values of $v$. In order to implement the IR cut-off, we need to understand how $f(v)$ in (\ref{bmans})  behaves for large values of $v$. The single cusp solution with boundary conditions at $r=0$ is
\begin{align}
\label{singleeasy}
T(u,v)&=e^v \cosh u,&X(u,v)=e^v \sinh u,&Z(u,v)&=\sqrt{2}e^v
\end{align}

As far from the cusp we expect the two solutions (with boundary conditions at $r=r_c$ and $r=0$) to look alike, we conclude that $f(v) \approx r_c e^v$ for large values of $v$.  In order to compute the area, we then need to integrate $e^{\hat \alpha}$ in the range in which $0<T(u,v)<t_c$. The second constraint implies $e^v \cosh u<t_c/r_c$, hence
\begin{equation}
A=\int du\, dv\, e^{\hat \alpha}=2 \int_0^{\log t_c} \tanh^2\left(\frac{v}{\sqrt{2}}\right) \text{arcosh} \left(\frac{t_c}{e^v} \right) = \log^2 (2 t_c)+\ldots
\end{equation}

where we have suppressed the $r_c$, as all depends on the ratio $t_c/r_c$. In order to compute the leading piece of the above integral, we have set $\tanh^2 \rightarrow 1$, which we can do if we assume that the contribution from the region $v \approx 0$ is small.

Expressing the leading contribution to the area as $A = \frac{1}{4} \log^2 {\frac{t_c^2}{r_c^2}}+\ldots$
(we have reinstated $r_c$) we see that the overall factor is exactly the same as the factor obtained in the first reference in \cite{AWLreviews} , so the value of the cusp anomalous dimension computed with this regularisation agrees with the well known result ($\sqrt{\lambda}/\pi$).

In principle one could compute the collinear anomalous dimension, characterising sub-leading IR divergences, from this solution. It is not clear whether the result will agree with the one obtained in the first reference in \cite{AWLreviews}, since we will have additional contributions that were not taken into account properly there. On the other hand, one expects the argument leading to the dual conformal ward identity presented in the first reference in \cite{AWLreviews} to go through, also when considering the ``correct'' solution.

\section{The infinitesimal form of the dual conformal generators}
\label{infconf}

The conformal group generators in dimension $d>2$ are given by \footnote{A factor of $(-i)$ was removed from all generators.}
\begin{align}\label{confgenerators}
D &=  x^{\mu} \partial_{\mu} \,,&M_{\mu \nu} &= -x_\mu \partial_\nu + x_\nu \partial_{\mu} \,, \\
P_{\mu} &= \partial_{\mu} \,  &K_{\mu} &= 2 x_{\mu}x^{\nu} \partial_{\nu} - x^2 \partial_{\mu}\,,
\end{align}
where $\partial_{\mu}:=\frac{\partial}{\partial x^{\mu}}$.
We recall that special conformal transformations can be obtained by doing an inversion, followed by a translation, and another inversion.
Since our integrals only have four-dimensional translation symmetry, only the corresponding four components
of the five-dimensional $\hat{K}^{M}$ will be symmetries of the integral.
Starting with (\ref{confgenerators}) in five dimensions and using $\hat{x}^{M} = (x^{\mu},m)$ we find
\begin{align}
\hat{D} &= x^{\mu} \partial_{\mu} + m \partial_{m}\,,\\
\hat{K}_{\mu} &= 2 x_{\mu} (x^{\nu} \partial_{\nu} + m \partial_{m}) - (x^2 +m^2) \partial_{\mu} \,,
\end{align}
where $\partial_{m} :=\frac{\partial}{\partial m}$.
In four dimensions, $(x_{i} -x_{j})^2$ is covariant under conformal boosts,
\begin{equation}
K^{\mu} (x_{i} -x_{j})^2 = 2 (x_{i}+x_{j})^{\mu}\, (x_{i}-x_{j})^2\,.
\end{equation}
In our case, the latter equation generalises to
\begin{equation}\label{covariance1}
\hat{K}^{\mu} (\hat{x}_{i} -\hat{x}_{j})^2 = \hat{K}^{\mu} \left[  (x_{i} -x_{j})^2 + (m_{i}-m_{j})^2 \right] =
 2 (x_{i}+x_{j})^{\mu}\, \left[  (x_{i} -x_{j})^2 + (m_{i}-m_{j})^2 \right] \,.
\end{equation}
Similarly, we have
\begin{equation}\label{covariance2}
\hat{K}^{\mu} m_{i} m_{j} = 2 (x_{i}+x_{j})^{\mu}\, m_{i} m_{j}\,.
\end{equation}
{}From (\ref{covariance1}) and (\ref{covariance2}) we see that
\begin{equation}
 \hat{K}^{\mu} \, \frac{m_{i} m_{j}} {\hat{x}_{ij}^2} =0\,.
\end{equation}
Note also that we have $\hat{K}^\mu O(m) = O(m)$, so that
the small $m$ expansion commutes with $\hat{K}^{\mu}$.

\section{\texorpdfstring{$AdS_{5}$}{AdS5} isometries and dual conformal symmetry generators}
\label{AdS5app}

Here we explicitly derive the form of the dual conformal symmetry generators acting
in the bulk of $AdS_{5}$.
We can define $AdS_{5}$ in embedding coordinates through the equation
\begin{equation}
- Y_{-1}^2 - Y_{0}^2 + Y_{1}^2 + Y_{2}^2 + Y_{3}^2 + Y_{4}^2 = -R^2\,,
\end{equation}
and we will set $R=1$ for simplicity.
We expect the classical string action to have an $SO(2,4)$ symmetry, whose infinitesimal generators are given by
\begin{equation}\label{rotations}
J_{MN} = Y_{M} \frac{\partial}{\partial Y^N} -Y_{N} \frac{\partial}{\partial Y^M}\,,
\end{equation}
where $\frac{\partial}{\partial Y^M} Y_{N} = \eta_{MN}$ and $\eta_{MN} = {\rm diag}(-1,-1,1,1,1,1)$.
On the other hand, we can define Poincar\'e coordinates
\begin{align}\label{poincare}
Y^{\mu} &= \frac{x^{\mu}}{r} \,,& Y_{-1} + Y_{4} &=\frac{1}{r} \,,& Y_{-1} - Y_{4} &=\frac{r^2 +x_{\mu} x^{\mu}}{r}\,,
\end{align}
where the $SO(1,3)$ indices $\mu = 0,1,2,3$ are raised and lowered using $\eta_{\mu\nu}$.
Now we can act with the generators (\ref{rotations}) on the equations given in (\ref{poincare}) and find
the action of the symmetry generators when acting on the Poincar\'e coordinates.
We find
\begin{align}
J_{-1,4}  &= r \partial_{r} + x^{\mu} \partial_{\mu} = \hat D\,, \\
J_{4,\mu}-J_{-1,\mu} &= \partial_{\mu}=\hat P_{\mu} \,,\\
J_{4,\mu} + J_{-1,\mu} &= 2 x_{\mu} ( x_{\nu} \partial^{\nu} + r \partial_{r}) - (x^2 +r^2)\partial_{\mu}
=\hat K_{\mu} \,,
\end{align}
and $SO(1,3)$ rotations $J^{\mu \nu}$ of course.
$\hat K_{\mu}=J_{4,\mu} + J_{-1,\mu}$ is precisely the conformal generator studied in section 2.\\

\end{document}